\documentclass[10pt,twocolumn,a4paper]{article}

\usepackage{hyperref}
\usepackage{amssymb,amsmath}
\usepackage{gensymb}
\usepackage[usenames]{color}
\usepackage{multirow}
\usepackage{txfonts}
\usepackage[percent]{overpic}
\usepackage{overpic}
\usepackage{bbold}
\usepackage{slashed} 
\usepackage{bm}
\usepackage{mathrsfs}
\usepackage{booktabs}
\usepackage{url}

\usepackage{titlesec}

\titleformat*{\section}{\normalfont\fontfamily{phv}\fontsize{12}{17}\selectfont}
\titleformat*{\subsection}{\normalfont\fontfamily{phv}\fontsize{10}{17}\selectfont}
\titleformat*{\subsubsection}{\normalfont\fontfamily{phv}\fontsize{10}{17}\selectfont}

\usepackage{amssymb,amsmath}
\usepackage{graphicx}
\usepackage[usenames]{color}

\newcommand{\bea}{\begin{eqnarray}}
\newcommand{\eea}{\end{eqnarray}}
\usepackage[percent]{overpic}
\usepackage{overpic}
\usepackage{natbib}

\begin{document}

\title{A magnetospheric dichotomy for pulsars with extreme inclinations} 
   \author{Fan Zhang \\
 		   \small{Gravitational Wave and Cosmology Laboratory, Department of Astronomy, Beijing Normal University, Beijing 100875, China} \\	  \small{Advanced Institute of Natural Sciences, Beijing Normal University at Zhuhai 519087, China}
          }
   \date{\small{\today}}

\twocolumn[
  \begin{@twocolumnfalse}
    \maketitle
    \begin{abstract}
{ In this work, we expand on a comment by Lyne et al (2017), that intermittent pulsars tend to congregate near a stripe in the logarithmic period versus period-derivative diagram. Such a stripe represents a small range of polar cap electric potential. Also taking into account the fact (already apparent in their Fig.~7, but not explicitly stated there) that high-fraction nulling pulsars also tend to reside within this and an additional stripe, we make the observation that the two stripes further match the ``death lines'' for double and single-pole interpulses, associated with nearly orthogonal and aligned rotators respectively. }
These extreme inclinations are known to suffer from pair production deficiencies, so we propose to explain intermittency and high-fraction nulling by reinvigorating some older quiescent (no pulsar wind or radio emission) ``electrosphere'' solutions. Specifically, as the polar potential drops below the two threshold bands (i.e., the two stripes), corresponding to the aligned and orthogonal rotators, their respective magnetospheres transition from being of the active pair-production-sustained type into becoming the electrospheres, in which charges are only lifted from the star. The borderline cases sitting in the gap outside of the stable regime of either case manifest as high-fraction nullers. Hall evolution of the magnetic field inside orthogonally rotating neutron stars can furthermore drive secular regime changes, resulting in intermittent pulsars. 
\vspace{1mm}
    \end{abstract}
  \end{@twocolumnfalse}
  ]

\section{Introduction}
Some pulsars show systematic long term variations in their emission behaviors, in the form of intermittency \cite{2006Sci...312..549K,2012ApJ...746...63C,2012ApJ...758..141L,2017ApJ...834...72L}, whereby the emission ceases for a significant period of time before turning back on again. Although rather preliminary in terms of statistics due to the small number of such systems so far observed, it has been noted by \cite{2017ApJ...834...72L} that these pulsars appear to congregate along a straight stripe in the period versus period-derivative or $\log_{10}P-\log_{10}\dot{P}$ diagram (see Fig.~\ref{fig:WithLabel}). We also point out that the a grouping of high-fraction nullers (we distinguish them from intermittent pulsars by their much shorter nulling periods -- comparable in order of magnitudes to pulse intervals; see also Sec.~\ref{sec:Discussion} for their difference with low-fraction nullers) around that stripe is also quite discernible, with furthermore, the remaining high fraction nullers seen (with significant scatter) to congregate near another similar stripe. It is noted by \cite{2017ApJ...834...72L} that such lines/stripes are likely significant because they correspond to constant potential drops across polar caps (abbreviated as $\mathscr{P}$ below), in the rudimentary vacuum dipole model of the pulsar magnetosphere (summarized in the first section of the appendix), which also serves as a baseline parameter for constructing more sophisticated models (summarized in the latter two sections of the appendix).

Zooming further out to encompass all pulsars, it is particularly noteworthy that not all pulsars near the proposed lines share such variability, so the variable population must correspond to special configurations possessing particular but not abnormal (otherwise the pulsars likely won't behave normally in their ``on'' states) parameters. Recalling that those lines of constant $\mathscr{P}$ are meaningful only when comparing pulsars of similar inclinations (angle between the rotational and magnetic axes, see Sec.~\ref{sec:Lines} below), it is then natural to suspect that these pulsars are nearly (not precisely, thus some scattering about the lines are to be expected) orthogonal or aligned rotators, respectively for the two lines, since such special inclinations are known to be problematic for secondary pair plasma production \footnote{E.g., they have both been proposed as terminal quiescent states that pulsars age toward \cite{1970ApJ...159L..81D,1970ApJ...160L..11G,1993ppm..book.....B}; note though that whether pulsars actually evolve towards either is more subtle, depending on details of the stellar shape and higher magnetic multipoles \cite{1970ApJ...160L..11G,2014MNRAS.441.1879P}, but this is irrelevant to the present discussion -- we are only concerned about the present values of the inclination angles, regardless of whether the pulsars evolved there over time or were born into them.}. The corresponding magnetospheres could then plausibly begin to fall into comatose states when the electromotive force across the polar caps drop below critical values (signified by the two stripes), for which radio emission and pulsar wind start to become suppressed.

In this paper, we investigate this possible association of some variable pulsar populations with extreme inclinations, beginning with an assessment of the observational evidences in Sec.~\ref{sec:Stats}, followed by theoretical speculations on the underlying physics in Sec.~\ref{sec:Theory}, before concluding in Sec.~\ref{sec:Discussion}.

\section{Statistics} \label{sec:Stats}
\subsection{Clustering} \label{sec:Lines}

\begin{figure}[ptb]
  \centering
\begin{overpic}[width=0.99\columnwidth]{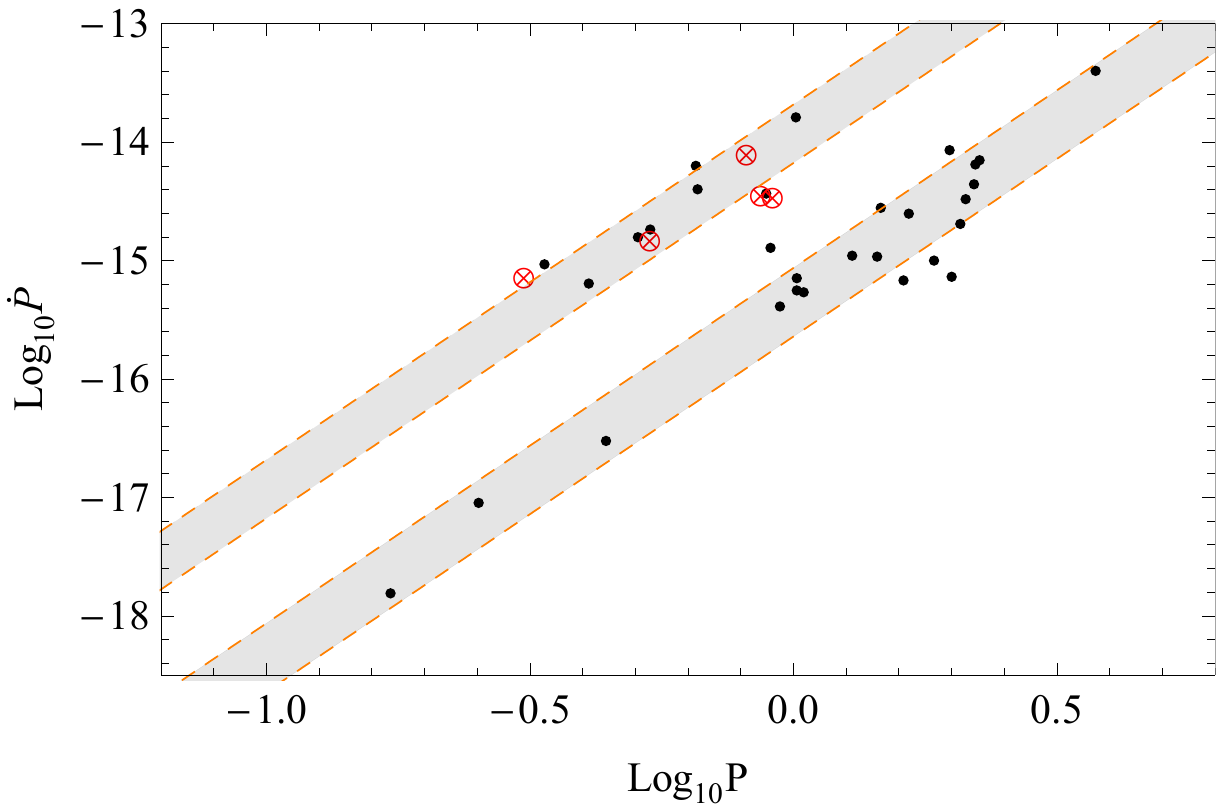}
\put(5,2){(a)}
\put(17,23){$\mathcal{O}$}
\put(20,13){$\mathcal{A}$}
\end{overpic}
\begin{overpic}[width=0.99\columnwidth]{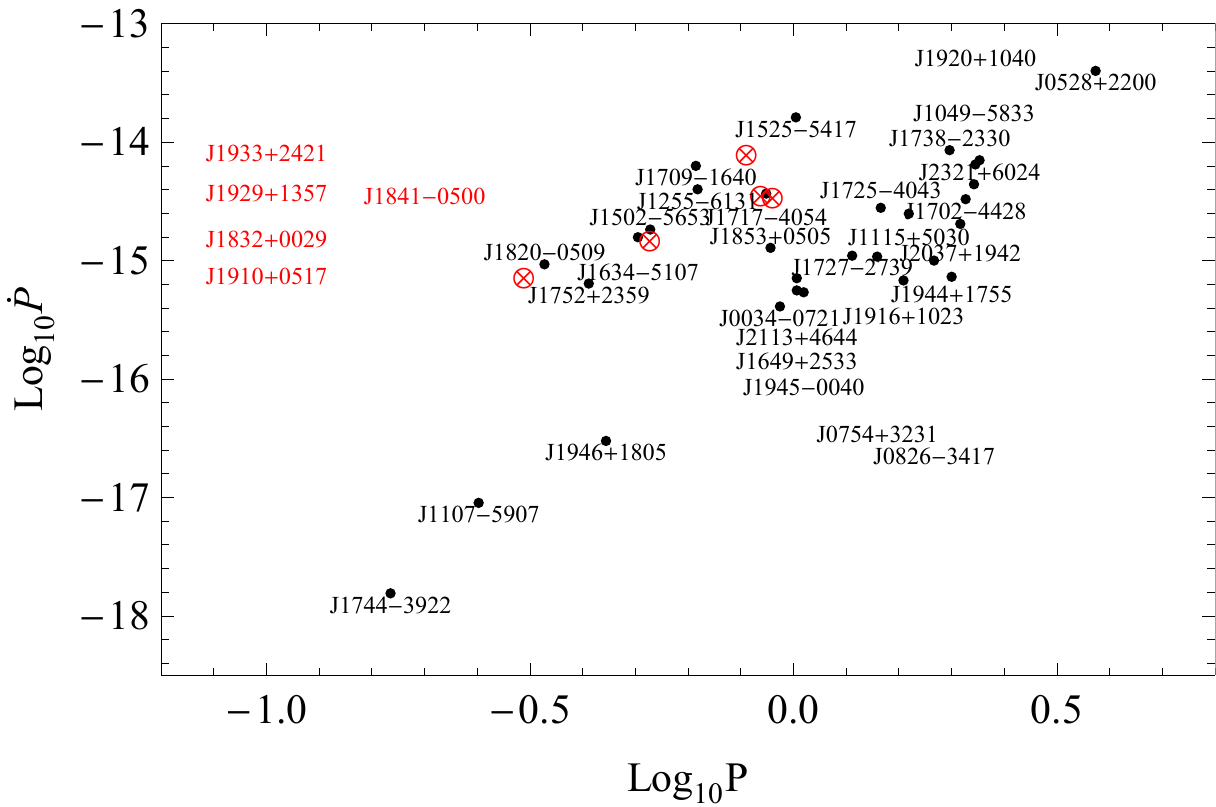}
\put(5,2){(b)}
\end{overpic}
  \caption{Placement of intermittent (red $\otimes$) and high-fraction nulling (black dots) pulsars on the $\log_{10}P-\log_{10}\dot{P}$ diagram.  
(a) Those nulling pulsars with $> 15\%$ null fraction [chosen due to B1133+16 apparently belonging to the low-fraction category given its non-simultaneous nulling across frequency bands \cite{2007A&A...462..257B}; we also caution that some nulling may be fake -- simply due to emission getting into a weaker mode below detection threshold, so a definitive null fraction threshold is unlikely meaningful], as well as the intermittent pulsars (essentially the same as Fig.~7 of \cite{2017ApJ...834...72L}, but with additional data points). The stripes correspond to the one standard deviation band of $\mathscr{P}$  associated with all the points clustered around it. 
(b) Same as (a), but with pulsar names displaced near the data points. Some names have to be offset to avoid overlapping. { The labels for the intermittent pulsars are placed at the same vertical height as the corresponding data point. If two labels are similar in height, their horizontal displacements reflect those of the data point. The labels for the nulling pulsars are arranged to coincide with the data points in their horizontal location, and when the labels become too crowded, their vertical height becomes the secondary discriminator that determines the corresponding data point.}
}
	\label{fig:WithLabel}
\end{figure}

\begin{figure}[tb]
  \centering
\begin{overpic}[width=0.99\columnwidth]{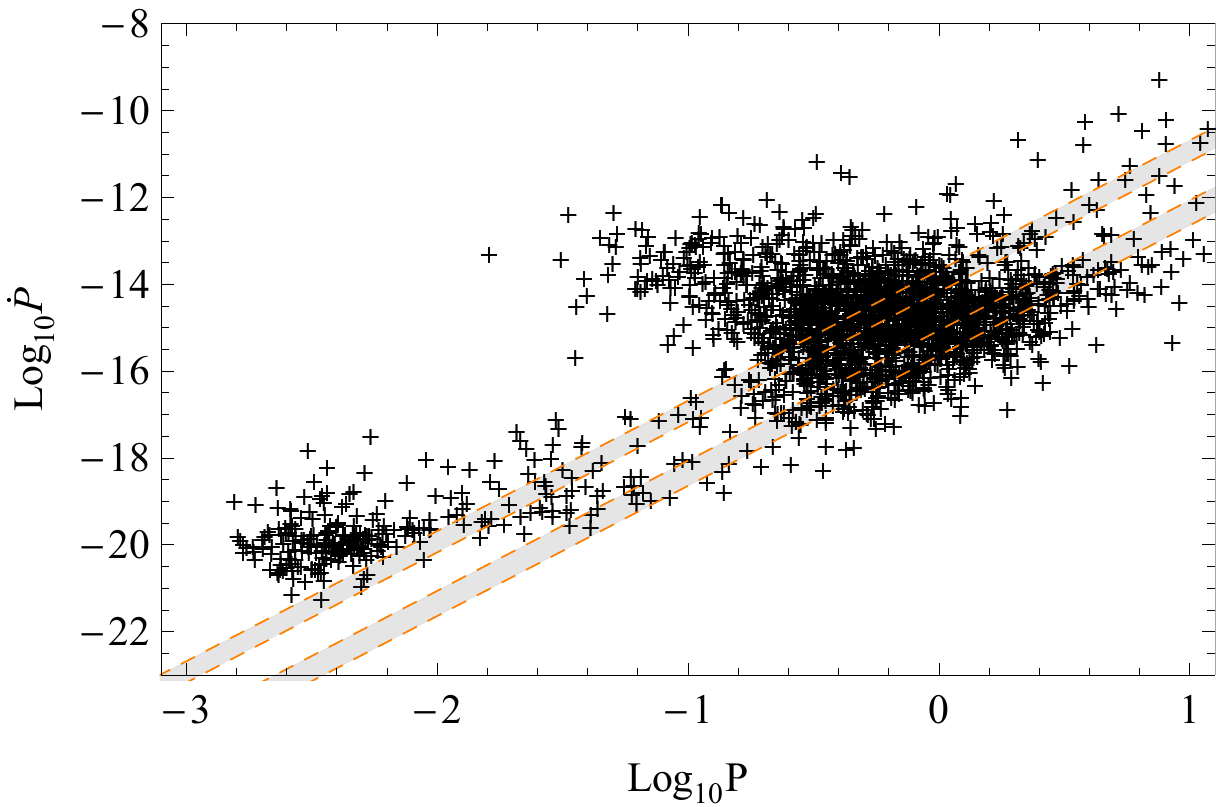}
\put(19,13){$\mathcal{O}$}
\put(30,13){$\mathcal{A}$}
\end{overpic}
  \caption{
For reference, the location of the stripes in relation to all known pulsars with a negative $\dot{P}$ (a few pulsars are measured to be spinning up, possibly due to the gravitational effects associated with their being inside globular clusters) is shown in this panel, data are from the ATNF catalogue. 
}
	\label{fig:AllPulsar}
\end{figure}

Fig.~\ref{fig:WithLabel} (a)\footnote{ The data for the intermittent pulsars are taken from \cite{2006Sci...312..549K,2012ApJ...746...63C,2012ApJ...758..141L,2017ApJ...834...72L}. The nulling data are taken from \cite{2007MNRAS.377.1383W} and references therein, namely \cite{1970ApJ...162..727H,1976MNRAS.176..249R,1979MNRAS.186P..39D,1983MNRAS.204..519L,1986ApJ...301..901R,1986AJ.....92..621W,1992ApJ...394..574B,2002AJ....123.1750L,2004ApJ...600..905L,2004MNRAS.355..147F,2005MNRAS.357..859R,2015MNRAS.449.1495Y}, and we also supplement the collection with more recent discoveries (including updates on null fractions to previously known ones) from \cite{2012MNRAS.423.1351B,2012MNRAS.424.1197G,2014MNRAS.439..221G,2018MNRAS.475.2375N}. We have also made use of the ATNF catalogue \cite{2005AJ....129.1993M,atnfweb}.}(b) are visually suggestive as to the clustering of intermittent pulsars and nullers in the $\log_{10}P-\log_{10}\dot{P}$ diagram, but eye-balling is not adequate, since e.g., the underlying overall pulsar population density (plotted in Fig.~\ref{fig:AllPulsar}\footnote{ One might notice in Fig.~\ref{fig:AllPulsar} that the $\mathcal{O}$ stripe appears to coincide with the death line for millisecond pulsars to the bottom left corner, a feature that may have relevance to the large fraction of such pulsars being identifiable as orthogonal rotators \cite{1996ASPC..105..231J,1993ApJ...408..179C}. We will however not pursuit this line of investigation further here, and focus on the regular population containing the high-fraction nullers and intermittent pulsars. }) may already be abnormally high in those $\mathscr{P}$ neighbourhoods, such that similar clustering patterns would arise if we just randomly drawing some samples out of the general population. We need a more quantitative gauge of how likely this feature arises purely by chance, especially since the data points are scarce and scattering is not negligible (recall that, to complicate matter, some scattering is unavoidable because the relevant pulsars inclination angles won't all be precisely $0$ or $\pi/2$). For example, we could compute the silhouette values for the members of these populations against the generic pulsar population displayed in Fig.~\ref{fig:AllPulsar}. Let $a_i$ denote the average distance in $\mathscr{P}$ (computed according to Eq.~\ref{eq:Line} below) of a sample point $i$ amongst e.g., nullers, to all other members of the same collection, and $b_i$ its average distance to all pulsars, then the silhouette
\bea
s_i = \frac{b_i-a_i}{\text{max}(a_i,b_i)}
\eea
satisfies $-1\leq s_i \leq 1$, and a value close to $1$ indicates vary tight clustering, i.e., $a_i \ll b_i$, while a negative value means the clustering is phantom, in that the sample point is more closely associated with the general population than with other points in its assigned collection.
The results are listed in Tb.~\ref{tb:silhouette}, and indeed all show large positive values. Note that we have also listed the silhouette values for stripe $\mathcal{O}$ nullers against the $\mathcal{A}$ grouping (i.e., $b_i$ are average distances to nullers on $\mathcal{A}$ rather than the general population), and vice versa. The fact all these values are positive indicates that we have not mis-assigned affiliation of points (reassign any nuller will result in negative values).    
 
\begin{table}[ptb]
  \begin{center}
    \caption{Silhouette values indicating the statistical significance of the clustering in $\mathscr{P}$ of the intermittent pulsars and high-fraction nullers. \\ }
    \label{tb:silhouette}
    \begin{tabular}{l|l|l}
      \toprule 
      \textbf{Pulsar } & \textbf{Silhouette } & \textbf{Silhouette }\\
      (Nuller Stripe $\mathcal{O}$) & vs. General Pop.& vs. Stripe $\mathcal{A}$ \\
      \midrule 
\text{J1525-5417}&0.84&0.88\\
\text{J1709-1640}&0.77&0.84\\
\text{J1255-6131}&0.85&0.89\\
\text{J1717-4054}&0.63&0.60\\
\text{J1502-5653}&0.85&0.88\\
\text{J1820-0509}&0.75&0.83\\
\text{J1634-5107}&0.85&0.88\\
\text{J1752+2359}&0.80&0.83\\
      \midrule 
      \text{Average}& 0.79 & 0.83 \\
      \midrule 
      \midrule 
      \textbf{Pulsar } & \textbf{Silhouette } & \textbf{Silhouette }\\
      (Nuller Stripe $\mathcal{A}$) & vs. General Pop.& vs. Stripe $\mathcal{O}$ \\
      \midrule 
\text{J2037+1942}&0.80&0.82\\
\text{J1725-4043}&0.76&0.74\\
\text{J1853+0505}&0.54&0.39\\
\text{J1944+1755}&0.67&0.71\\
\text{J1107-5907}&0.86&0.86\\
\text{J1049-5833}&0.86&0.87\\
\text{J1702-4428}&0.85&0.86\\
\text{J1727-2739}&0.86&0.86\\
\text{J1916+1023}&0.76&0.78\\
\text{J1920+1040}&0.85&0.85\\
\text{J0034-0721}&0.87&0.87\\
\text{J0528+2200}&0.80&0.79\\
\text{J0754+3231}&0.86&0.86\\
\text{J0826-3417}&0.75&0.78\\
\text{J1115+5030}&0.86&0.86\\
\text{J1649+2533}&0.86&0.86\\
\text{J1744-3922}&0.84&0.85\\
\text{J1945-0040}&0.87&0.87\\
\text{J1946+1805}&0.85&0.86\\
\text{J2113+4644}&0.83&0.82\\
\text{J2321+6024}&0.85&0.84\\
\text{J1738-2330}&0.70&0.65\\
      \midrule 
      \text{Average}& 0.81 & 0.80 \\
      \midrule 
      \midrule 
      \textbf{Pulsar } & \textbf{Silhouette } & \\
      (Intermittent) & vs. General Pop. & \\
      \midrule 
\text{J1832+0029}&0.76&\\
\text{J1841-0500}&0.66&\\
\text{J1910+0517}&0.59&\\
\text{J1929+1357}&0.71&\\
\text{J1933+2421}&0.72&\\
      \midrule 
      \text{Average}&0.69& \\
      \bottomrule 
      \end{tabular}
  \end{center}
\end{table}

\begin{table}[tb]
  \begin{center}
    \caption{The ratio of Monte Carlo simulated samples (with the same number of pulsars in each sample as in the observed population tagged by the first column) with an average silhouette value equal or above that of the actual observed population. { Note that sample size refers to the number of Monte Carlo simulations, not the number of pulsars in the mock up samples.}\\}
    \label{tb:MC}
    \begin{tabular}{l|l|l}
      \toprule 
      \textbf{Pulsar Population} & \textbf{Sample Size} & \textbf{Ratio $\geq$ Observed}\\
      \midrule 
\text{Intermittent}&$10^4$&$0.69\%$\\
&$10^5$&$0.65\%$\\
&$10^6$&$0.67\%$\\
&$10^7$&$0.67\%$\\
      \midrule 
\text{Nuller Stripe $\mathcal{O}$}&$10^4$&$0\%$\\
&$10^5$&$0.0020\%$\\
&$10^6$&$0.0018\%$\\
&$10^7$&$0.0020\%$\\
&$10^8$&$0.0020\%$\\
      \midrule 
\text{Nuller Stripe $\mathcal{A}$}&$10^4$&$0\%$\\
&$10^5$&$0\%$\\
&$10^6$&$0\%$\\
&$10^7$&$0\%$\\
&$10^8$&$0\%$\\
      \bottomrule 
      \end{tabular}
  \end{center}
\end{table}

This simple (thus robust) silhouette would already constitute a quantitative measure, but we must also correct for the influence of our small sample sizes. To achieve ``normalization'' (and also to provide a more intuitive interpretation of what the silhouette values imply), we can further compare the mean silhouette value $\bar{s}$ for a observational sample against its counterparts computed for the same-sized { mock up} samples drawn randomly via Monte Carlo simulations from the general pulsar population 
{ (i.e., 8 general pulsars are randomly picked to mock up a sample whose silhouette value can be compared with that of nullers in stripe $\mathcal{O}$; similarly 22 are plucked for comparison with nullers in stripe $\mathcal{A}$ and 5 for the intermittents)}. 
The simulations as such provide a distribution of $\bar{s}$ appropriate for this particular sample size, and where the physical ensemble lands within this distribution (specifically, what ratio of Monte Carlo outcomes end up with greater or equal $\bar{s}$ than the observational sample) tells us the likelihood of the stripe-like-grouping arising purely by chance -- i.e., the probability that the intermittent pulsars and high-fraction nullers have no special preference for $\mathscr{P}$, as compared to the general population.  

The simulations are trivial to carry out. One merely needs to draw random numbers from a uniform distribution over the index set of all pulsars, and then the pulsars whose numbers get called join the simulated ensemble. This is analytically equivalent (but numerically more efficient and accurate) to the more spelled-out procedure of extracting the joint distribution of pulsars over the $\log_{10}P-\log_{10}\dot{P}$ plane, collapsing it into a marginal distribution over $\mathscr{P}$ by integrating out the directions of constant $\mathscr{P}$ on the $\log_{10}P-\log_{10}\dot{P}$ plane, and then drawing random samples by inverting the resulting cumulative distribution function over a uniform distribution over the interval $[0,1]$. The results of our simulation are displayed in Tb.~\ref{tb:MC}, which show that very small percentages of samples have greater or equal $\bar{s}$ than the observed populations,  thus demonstrating that the observational clusterings are unlikely chance occurrences. Converting to the familiar Gaussian sigma terminology, the odds of the three clusters arising by chance are $\approx2.7\sigma$ for intermittent pulsars, $\approx4.3\sigma$ for nullers near $\mathcal{O}$ and $>5.7\sigma$ for nullers near $\mathcal{A}$ (up to the maximum $10^8$ samples that our computing resources permit, there has been no instance of equally tight or tighter clustering arising by chance, which sets an upper limit of $10^{-8}$, and thus an inequality). 
Note the progression of increasing number of sigmas is due to the differences in population sizes, matching the intuition that it should be more unlikely for larger populations to cluster by chance. This is what we meant by ``normalization by sample size'' earlier. 

It should also be noted that we have split the nullers into two sub-populations that are examined separately, an action that has thus far been based on our theoretical understanding that there exists of two classes of problematic magnetospheres -- the aligned and orthogonal rotators. However, if one prefers statistics extracted from unadulterated purely observational data, without introducing theoretical priors, then they perhaps should consider all nullers as a single collection, and count all Monte Carlo samples that can be subdivided into a small number of very tightly clustered subgroups as being equally good or superior. We expect this alternative approach to yield slightly less extreme number of sigmas, depending on how many sub-clusters we allow and how we average their silhouettes. This fact that there are many tunable knobs is a significant deficiency though. In particular, we don't see how theoretical priors can be prevented from sneaking back in, through whichever way we penalize larger numbers of sub-clusters (at the very least we need to decide on how many is too many, inevitably evoking a theoretical expectation of how many there should be). Failing this prevention would then defeat the purpose of introducing this more opaque and technically more complicated approach in the first place, so we do not adopt it here. Furthermore, as we now discuss in the next section, there in fact exists observational evidence for the two stripes being associated with aligned and orthogonal rotators respectively, thus our division of the high-fraction nuller population in two is not solely theoretical. 

\subsection{Inclination angles} \label{sec:Angles}

For a small number of pulsars, relatively precise measurements of the inclination angle $\alpha$, by fittings to polarization position angles or PPA (as per \cite{1969ApL.....3..225R}), is available. Near $\mathcal{O}$, \cite{2007ARep...51..477M} showed that the intermittent J1933+2421 is indeed a nearly orthogonal rotator. In addition, measurements for J1841-0500 had been done by \cite{2012ApJ...746...63C}, giving an impact parameter of $\beta \lesssim 3^{\circ}$. Although a measurement on $\alpha$ is not explicitly given in that work, we can infer from the very steepest slope of the PPA curve (the steepest part equals $\sin\alpha/\sin\beta$) that $\sin\alpha$ should be quite close to saturating out at $\sin\pi/2$. For nullers near line $\mathcal{A}$ on the other hand, explicit position angle studies by \cite{2010MNRAS.408...40K} ($\lesssim 2^{\circ}$ for J1944+1755) and \cite{2014MNRAS.442.2519Y} (for J1107-5907) have also confirmed that the inclination angle $\alpha$ is indeed very small. The high-fraction ($70\%$) nuller J0826-3417 likewise has a very small $\alpha$  since the emission covers the entire pulse longitude (we are always inside the emission cone) \cite{1979MNRAS.186P..39D}. 

\begin{figure}[tb]
  \centering
\begin{overpic}[width=0.99\columnwidth]{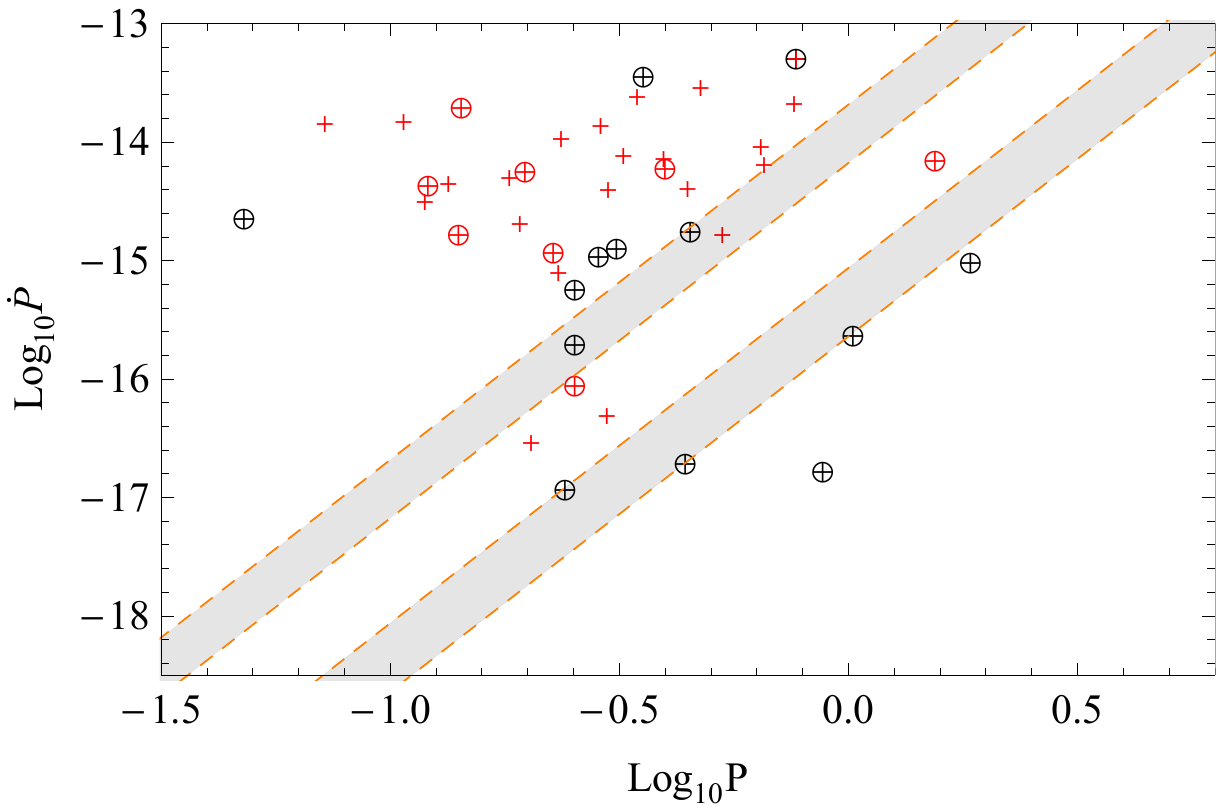}
\put(14,12){$\mathcal{O}$}
\put(31,12){$\mathcal{A}$}
\end{overpic}
  \caption{
Pulsars (not necessarily nulling or intermittent) exhibiting interpulses of both double-pole (red crosses; almost orthogonal rotators) and single-pole (black circled crosses; almost aligned rotators) as compiled by \cite{2011MNRAS.414.1314M}  (with data originally from \cite{2004A&A...428..943K,1993ApJS...88..529T,2008MNRAS.391.1210W,1998MNRAS.297...28D,2002MNRAS.335..275M,2003MNRAS.342.1299K,2004MNRAS.352.1439H,2006MNRAS.372..777L,2001MNRAS.328...17M,2009A&A...498..223J,2009MNRAS.395..837K,2010MNRAS.402..745K,2009ApJ...705....1C}), as well as \cite{2013ARep...57..833M}. The red circled crosses are pulsars for which \cite{2011MNRAS.414.1314M} and \cite{2013ARep...57..833M} made opposite single or double-pole assignments. 
}
	\label{fig:AngleStats}
\end{figure}

Another supporting evidence for associating $\mathcal{O}$ and $\mathcal{A}$ stripes with inclinations is that since they are essentially the ``death lines'' for orthogonal and aligned rotators within our proposal, pulsars (not necessarily nulling or intermittent) that exhibit robust characteristics confirming their extremal inclinations should exist only to the upper-left regions of these stripes. One particular feature of orthogonal and aligned rotators is that they often (but not always, depending on the detection threshold among others, but for our purpose we only need this effect to be sufficient and not necessary) exhibit interpulses, of the double and single-pole variety respectively \cite{2011MNRAS.414.1314M}. With orthogonal rotators, it is because the emission cones from both poles could sweep across Earth, so interpulses sit roughly half-way in-between ``regular'' pulses. With nearly aligned rotators, the ``regular pulse''-interpulse pair would be due to the Earth traversing the two walls of a hollow conic emission beam \cite{1968Natur.218..934R}, usually with a bridge connecting the two (see e.g., \cite{1979MNRAS.186P..39D} for J0826-3417). This is essentially an extreme version of the double-poled pulse profile, whereby the two peaks are extra-widely separated due to the small $\alpha$, to the extent that they are (mis-)identified as separate pulses. 
Alternatively, the single-pole interpulses can be due to the line of sight being always within the beam (possible for nearly aligned rotators), and threading through two nested cones or a core-conal duo \cite{1983A&A...127..267G,1985ApJ...299..154G}. With either interpretation, single-pole interpulses flag nearly-aligned rotators. 
In Fig.~\ref{fig:AngleStats}\footnote{  The peculiar fact that the double-pole pulsars are located much further to the top-left has already been noticed by the cited references, although a connection with nullers were not made.   
Unfortunately, these real-world pulsars are not perfectly aligned or orthogonal, so the death lines won't be sharp (some overshooting is to be expected; for the same reason, data point scattering in Fig.~\ref{fig:WithLabel} is unavoidable), and their inclination angles are furthermore not in fact precisely known, so a quantitative assessment of the statistical significance of this feature is difficult.}, we plot the $\log_{10}P-\log_{10}\dot{P}$ locations of the pulsars exhibiting double and single-pole interpulses (as compiled by \cite{2011MNRAS.414.1314M}), and they indeed appear to reside mostly to the upper left of the relevant stripes.  

Once again, we could assess this observation more quantitatively, by defining a quantity that we term the ``penumbra'', that measures the amount of spillover beyond the relevant death lines: 
\bea
p_{\mathcal{D}}=\frac{1}{n}\sum_i \text{max}\bigg(\mathscr{P}_{\mathcal{D}}-\mathscr{P}_i,0\bigg)\,, 
\eea
where $i$ indexes the single or double-pole population, $n$ is their population size, $\mathcal{D}=\mathcal{O}$ or $\mathcal{A}$, and $\mathscr{P}_{\mathcal{D}}$ is associated with the center of the relevant stripe (average of the $\mathscr{P}$ values for the intermittent and/or nulling pulsars clustered around the stripe, see caption to Fig.~\ref{fig:WithLabel}). For any prescribed death line, the penumbra vanishes when all data points lie to the upper-left of it. It does not however, by itself, tells us whether the death line is appropriate, in the sense that one could obviously place it very far to the bottom-right of the $\log_{10}P-\log_{10}\dot{P}$ diagram, so that the interpulse (and indeed all) pulsars lie to its top-left trivially. Such a situation conjures little statistical significance.  

\begin{table}[tb]
  \begin{center}
    \caption{The Monte Carlo simulation results for the penumbra values. The qualification ``Uncertain to Single'' means the circled red crosses of Fig.~\ref{fig:AngleStats} are grouped with the single-poled population. { Sample size here refers to the number of Monte Carlo simulations carried out.} \\
    }
    \label{tb:MCInterpulse}
    \begin{tabular}{l|l|l}
      \toprule 
      \textbf{Pulsar Population} & \textbf{Sample Size} & 
     \textbf{Ratio $\leq$ Observed}\\
     \text{Uncertain to Single} & & \\
      \midrule 
\text{Single-poled}&$10^4$&$68\%$\\
&$10^5$&$68\%$\\
&$10^6$&$68\%$\\
&$10^7$&$68\%$\\
      \midrule 
\text{Double-poled}&$10^4$&$0\%$\\
&$10^5$&$0.0030\%$\\
&$10^6$&$0.0015\%$\\
&$10^7$&$0.0012\%$\\
&$10^8$&$0.0012\%$\\
      \midrule 
      \midrule 
      \textbf{Pulsar Population} & \textbf{Sample Size} & \textbf{Ratio $\leq$ Observed}\\
     \text{Uncertain to Double} & & \\
      \midrule 
\text{Single-poled}&$10^4$&$88\%$\\
&$10^5$&$87\%$\\
&$10^6$&$87\%$\\
&$10^7$&$87\%$\\
      \midrule 
\text{Double-poled}&$10^4$&$0\%$\\
&$10^5$&$0\%$\\
&$10^6$&$0.00030\%$\\
&$10^7$&$0.00023\%$\\
&$10^8$&$0.00021\%$\\
      \bottomrule 
      \end{tabular}
  \end{center}
\end{table}

To handle this issue and to provide better intuition, we can once again evoke the Monte Carlo simulation, and record what percentage of simulated sample populations, of the same size as the single- or double-poled pulsars, possess smaller or equal penumbras than the actual observed population. Or in other words, to what likelihood a population of this size, residing mostly to the top-left of the death lines, would arise purely by chance. This gauge also penalizes over-generous death lines, since pushing them too far towards the bottom-right corner would lead to greater portions of simulated samples possessing equal or smaller penumbras, reducing (as should be) the statistical significance readings. This in fact causes trouble with the $\mathcal{A}$ stripe, which is unfortunately very close to the overall pulsar death line (see Fig.~\ref{fig:AllPulsar}), so the probability of the single-pole population residing mostly to its top-left purely by chance is quite high (see the ``Single-Poled'' rows in Tb.~\ref{tb:MCInterpulse}), and we cannot assert this stripe's role as a death line for the aligned rotators based on interpulse population statistics. Beyond the available $\alpha$ measurements for the few nullers discussed at the beginning of this section, we will have to base our confidence in this hypothesis mostly on its agreement with theoretical predictions. 
Note though, this feature of $\mathcal{A}$ sitting far to the bottom right does not translate into aligned rotators being easy to turn on, and is instead simply the result of small $\alpha$ causing large $\Delta \Phi^{\rm vac}$ values to map into small $\mathscr{P}$ (i.e., for similar stripe placements, much larger potential drops are demanded for aligned rotators than for inclined rotators, to which the overall death line refer), see Eq.~\eqref{eq:LineSmallAngle} below.
Fortunately, the $\mathcal{O}$ stripe sits much further up the $\mathscr{P}$ spectrum, so meaningful statistics ($\approx 4.4\sigma$ or $\approx 5.6\sigma$ depending on ambiguous population grouping, see Tb.~\ref{tb:MCInterpulse}) can be obtained for it, validating $\mathcal{O}$ as a likely death line for orthogonal rotators. 

So far our conclusions have been based on observational statistics, but theoretical arguments can be brought in to further enhance our confidence, that $\mathcal{O}$ is associated with orthogonal rotators. Specifically, with intermittent pulsars, it is quite natural to expect that the ``on'' and ``off'' states correspond to two modes of the pulsar magnetosphere, one with active pulsar winds and another with only vacuum-like dipole radiation. The simplest baseline estimate for the ratio of spin-down rates between them is \cite{2006ApJ...648L..51S}
\bea \label{eq:Ratio}
\frac{3}{2}\frac{1+\sin^2\alpha}{\sin^2\alpha}\,,
\eea 
which { while unfortunately} always overshoots the { observationally measured values that sit in the range of} $1.5-2.5$ \cite{2007Ap&SS.308..569B,2007MNRAS.377.1663G}. { Nevertheless, this expression} attains its minimal value {(closest to the observed values)} of $3$ at $\alpha=\pi/2$ (this ratio diverges for aligned or anti-aligned rotators in contrast), meaning it would be much easier to tweak the model (specifically in our case, the ``off'' state are electrospheres rather than actual vacuum, while the ``on'' state deficiencies of orthogonal rotators that we discuss in Sec.~\ref{sec:Active} have not been included in Eq.~\ref{eq:Ratio}) to match observations if intermittent pulsars are orthogonal rotators.   

\begin{figure}[tb]
  \centering
\begin{overpic}[width=0.75\columnwidth]{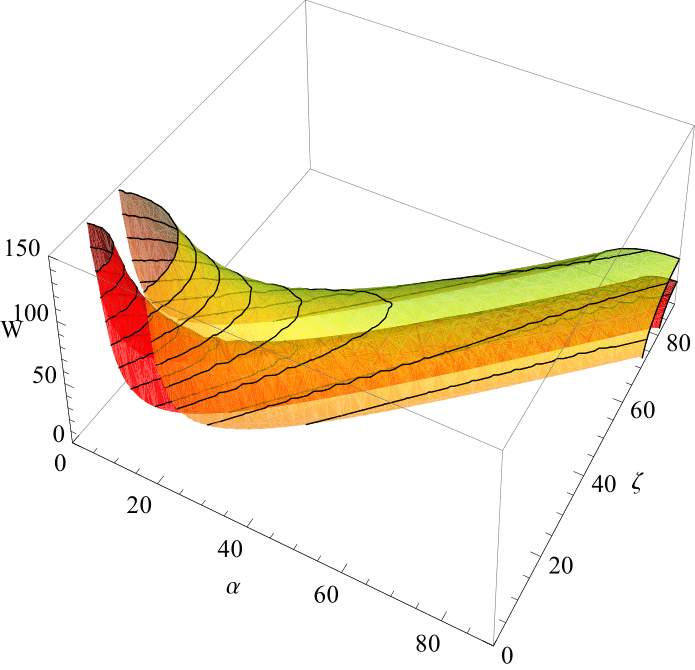}
\put(5,2){(a)}
\end{overpic}
\begin{overpic}[width=0.75\columnwidth]{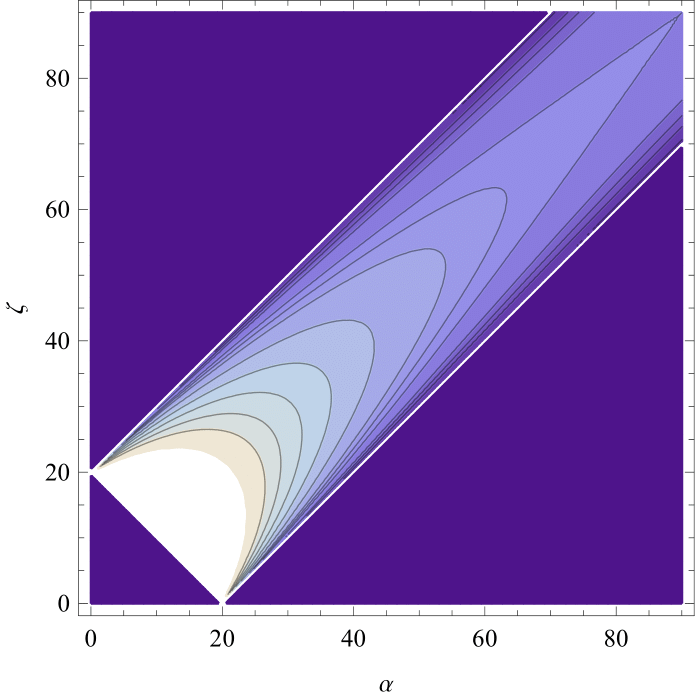}
\put(5,2){(b)}
\end{overpic}
  \caption{(a) The pulse width $\mathcal{W}$ as a function of the inclination angle $\alpha$ and viewing impact angle $\zeta$ { (see above Eq.~\ref{eq:W2} for its definition and below Eq.~\ref{eq:W1} for the definition of $\rho$)}. All values are in degrees. The semi-transparent surfaces correspond to $\rho=20^{\circ}$ (yellow) and $\rho=10^{\circ}$ (red). The Contours of constant $\mathcal{W}$ span wide $\alpha$ ranges, suggesting that $\alpha$ estimates based off of pulse width can not be very precise. (b) The contours of $\mathcal{W}$ on the $(\alpha,\zeta)$ plane, for the case of $\rho=20^{\circ}$.}
	\label{fig:WPlot}
\end{figure}

Lastly, for completeness, we note that due to the scarcity of precise inclination angle estimates based on position angle sweeps, indicative estimates have frequently been produced in literature using pulse width measurements, with \cite{Gil1981}
\bea \label{eq:W1}
\mathcal{W}(\alpha,\rho,\beta)=4 \arcsin\left(\sqrt{\frac{\sin\frac{\rho+\beta}{2}\sin\frac{\rho-\beta}{2}}{\sin\alpha\sin(\alpha+\beta)}}\right)\,,
\eea
where the emission { beam width $\rho$ (recall that $\beta$ is the impact parameter)} is commensurate with the emission level (e.g., $50\%$ or $10\%$) relevant for the pulse width $\mathcal{W}$. This method suffers from great uncertainty though. The sensitive dependence on $\beta$ in particular, causes considerable variation in the estimate of $\alpha$ from $\mathcal{W}$. Most crudely but conveniently, one may take $\beta =0$, and obtain the commonly evoked approximation $\mathcal{W}\approx 2\rho/\cos\alpha$ \cite{2011MNRAS.414.1314M}. A more complex empirical rule have also been proposed \cite{1990ApJ...352..247R} based on the existence of a Lower Boundary Line, where 
\bea \label{eq:Crude2}
\alpha \approx \arcsin\left( \frac{2.45^{\circ} P^{-0.5}}{\mathcal{W}_{\text{core}}}\right)\,,
\eea
with $\mathcal{W}_{\text{core}}$ being pulse width at $50\%$ strength of the core component in degrees. Even Eq.~\eqref{eq:Crude2} is too broad-stroked though (only useful when averaging over a large population of pulsars), since e.g., for J1944+1755 the position angle analysis gives $\alpha \lesssim 2^{\circ}$ \cite{2010MNRAS.408...40K}, while Eq.~\eqref{eq:Crude2} gives $14^{\circ}$ (assuming the whole width is core). To avoid misleading conclusions then, one must consider a range of $\alpha$ values achieved by allowing $\beta$ to vary. To this end, it is more convenient to replace $\beta$ with $\zeta=\alpha+\beta$, i.e., the angle between the spin axis and the observer, yielding \cite{1977puls.book.....M}
\bea \label{eq:W2}
\mathcal{W}(\alpha,\rho,\zeta)=2 \arccos\left( \frac{\cos\rho-\cos\alpha\cos\zeta}{\sin\alpha \sin\zeta}\right)\,,
\eea
which is relevant only when $|\alpha-\rho| \leq \zeta \leq \alpha+\rho$, ensuring that the pulsar is seen on Earth as pulsed emissions \cite{2010MNRAS.402.1317Y}. We plot in Fig.~\ref{fig:WPlot} the $\mathcal{W}$ values as functions of $\alpha$ and $\zeta$ for a couple of example choices $\rho=10^{\circ}$ and $20^{\circ}$. The contours of constant $\mathcal{W}$ shows that $\alpha$ can span vast ranges, so $\alpha$ derived from such considerations tend not to be very constraining. In particular, the contour plot of Fig.~\ref{fig:WPlot}(b) shows that $\alpha=0$ is always an admissible limiting solution in the $\zeta \rightarrow \rho$ limit, while maximum $\alpha$ value occurs close to the diagonal $\alpha=\zeta$ line. Therefore, for any width $\mathcal{W}_0$ corresponding to a particular pulsar, the value $\alpha_{\rm diag}$ as given by $\mathcal{W}(\alpha_{\rm diag},\rho,\alpha_{\rm diag})=\mathcal{W}_0$ provides an indication of the maximum $\alpha$ achievable for that pulsar. 

Note also that in addition to $\beta$ or equivalently $\zeta$, $\rho$ is also variable. One can in principal take the empirical $\rho_{\rm 10\%} \approx 4.9^{\circ}P^{-1/2}$ to $6.3^{\circ} P^{-1/2}$ \cite{1993ApJ...405..285R,1993A&A...272..268G,1994A&AS..107..515K} for the general pulsar population, but since we are examining geometrically special aligned and orthogonal rotators here, their beam widths may not exactly follow this rule. Furthermore, for many of the high-fraction nullers depicted in Fig.~\ref{fig:WithLabel}, it turns out that there isn't a solution for $\alpha_{\rm diag}$, since the corresponding contour does not intersect the diagonal line in Fig.~\ref{fig:WPlot}(b) for the given $\rho$ (i.e., the contours corresponding to those crowding near the boundary of the rectangular region; equivalently the ones at lowest $\mathcal{W}$ values in Fig.~\ref{fig:WPlot}(a)), meaning that there is no constraints on $\alpha$ at all. In summary, constraints on the inclination angle from pulse width is not particularly informative in our context, and we shall not pursue this avenue further. 

\section{Proposed explanation} \label{sec:Theory}
\subsection{Magnetospheric dichotomy}
Finer details within the $\log_{10}P-\log_{10}\dot{P}$ diagram can clue us in on interesting features of pulsar magnetospheres, so it is worthwhile coming up with a theoretical model that explains the stripes. {In other words, it is interesting to see} what surprising structural properties of the magnetosphere they may be hinting at. We propose that charge-abundant active magnetospheres and charge-starved quiescent electrospheres { (summarized in the latter two sections of the appendix)}
describe the live and dead pulsars that are separated by the $\mathcal{O}$ and $\mathcal{A}$ stripes. { These stripes thus} serve as the death lines for (nearly) orthogonal and aligned rotators, respectively.  

The discriminating factor determining which of these magnetospheric states is realized is whether pair cascade occurs to provide an abundance of charged particles. Typically, cascading behavior tends to emerge only when a threshold is crossed in the parameter space. It is quite plausible that aligned and orthogonal rotators, that have been known to suffer from different deficiencies with regard to pair production, will cross below said threshold once the electromotive force across the polar caps drops to their respective critical levels, as signified by the stripes $\mathcal{A}$ or $\mathcal{O}$. Indeed, numerical work by e.g., \cite{2013MNRAS.429...20T} (for 1-D) and \cite{2014ApJ...795L..22C,2015ApJ...815L..19P} (for aligned rotators) had indicated that the pulsar magnetospheres would revert to the charge-separated electrospheres when the pair production threshold exceeds the full polar cap potential. 

In other words, for pulsars with extreme inclinations and residing to the lower right of the relevant stripe, their $e^{\pm}$ cascade shuts down, leaving only those charges lifted directly out of the stars available to short out the parallel (to the magnetic field) electric field $E_{\parallel}$ and achieve force-freeness\footnote{Arising from negligible particle inertia as compared to electromagnetic energy density, so Lorentz forces must vanish or else infinite accelerations result. This is only relevant where there are charged particles, and not required in vacuum gaps. Note also that some literature took ${ E}\cdot { B}=0$ to mean force-free, even though it is only a necessary and not sufficient condition.} in the charge regions. This in turn reduces plasma multiplicity, driving the magnetospheres into the quiescent charge-separated electrospheres that those earlier studies had seen\footnote{These studies suitably assumed that the star is the sole source of charges. Heuristically, \cite{2001RMxAC..10..168M} explained that the charge-separated magnetospheres would naturally arise from turning up the magnetic field and thus the induced electric potential to strengths beyond the work function of the neutron star, so that the charges are gradually lifted out of the star but confined to nearby regions (because e.g., they cannot cross magnetic field lines easily, due to the strong magnetic field inducing synchrotron radiation that push the particles into fundamental Landau levels).}. This subsequently shuts down pulsar winds, since there are no current flows, as well as radio emission, which is assumed to emerge from near the separatrix current sheets (missing in electrospheres) in some models (see e.g., \cite{2018A&A...612A..24Z}), and otherwise modeled in neutral and not charge-separated (as in the electrosphere case) plasma in the rest (see e.g., \cite{2020PhRvL.124x5101P}).

In essence then, we propose that both sides of the aforementioned historical debate on the magnetospheric structure are in fact correct, with both types of magnetospheres realized in nature, reigning their respective realms (we will slight abuse terminology in the interest of brevity, and use ``pulsars'' to refer to both cases, even though the electrospheres give off no radio pulses). This paper is of course not the first to touch on such a bimodal possibility, e.g., \cite{2010HEAD...11.1621M,2002A&A...384..414P} motivated the study of electrospheres by noting that they would be useful if for some reason pair cascade ceases. The novelty of our work is mainly in the proposed role played by extreme inclinations.

There are also the more marginal cases where the nearly extreme inclination pulsars reside close to the relevant death lines, and hop back-and-forth between active and quiescent states. The magnetospheres for these pulsars evolve dynamically and thus should differ from either of the stationary cases described in the appendix. Absent full 3-D first-principle (including pair production microphysics) particle-in-cell numerical simulations, an accurate picture for such dynamical states is difficult to obtain (although already quite sophisticated simulations like those done by \cite{2012ApJ...746L..24L,2019arXiv191100059C} may have already captured many essential qualitative aspects; note that with the first reference, although their modeling is aimed at intermittent pulsars, the description could plausibly also be applied to the nulling cases), but we know such states can reside within the gap between the stability thresholds, $\mathscr{P}_{\mathcal{O}/\mathcal{A}}^a$ and $\mathscr{P}_{\mathcal{O}/\mathcal{A}}^e$ for the active and electrospheric magnetospheres respectively, where both those stationary configurations become unstable. 
Specifically, if 
\bea \label{eq:Stripe}
\mathscr{P}_{\mathcal{O}/\mathcal{A}}^a > \mathscr{P}_{\mathcal{O}/\mathcal{A}}^e\,,
\eea
and if the actual potential $\mathscr{P}_{\mathcal{O}/\mathcal{A}}$ for a nearly aligned or orthogonal rotator satisfies
\bea
\mathscr{P}_{\mathcal{O}/\mathcal{A}}^a > \mathscr{P}_{\mathcal{O}/\mathcal{A}}>\mathscr{P}_{\mathcal{O}/\mathcal{A}}^e\,,
\eea
then it cannot be steadily ``off'', nor can it remain consistently active. It would be stuck in a cycle of ``somewhat on'' and ``somewhat off'', manifesting as a high-fraction nuller. 

\subsection{Short circuiting the electrospheres} \label{sec:ShortCircuiting}
A physically significant feature of the electrospheres is that $E_{\parallel}$ is shorted out only within the charge clouds that are confined to regions close to the star (mostly inside the trap surfaces). All of the regions outside of the clouds are vacuum ``gaps'' whereby $E_{\parallel}$ is not shielded, and are thus possible sites of pair production, as has been highlighted by e.g., \cite{2002A&A...384..414P}. Such additional sources of charges are not included in the computations of the electrospheres, and could thus in principle invalidate them. However, the electrospheres should be stable against \emph{small amounts} of charge injections \cite{2001MNRAS.322..209S}, given that the direction of the electric field in the gaps is such that it transports the particles to the clouds of the same charges, thus maintain charge separation.

However, if these ``gap'' regions host strong enough $E_{\parallel}$ to drive runaway pair cascade, then we will witness copious flows of particles along certain magnetic field lines, esp.~along the would-be separatrix current sheets \cite{2019arXiv191100059C}, turning on global current circulation within the magnetosphere \cite{1997MNRAS.287..262S}, with current density reaching such intensity that the magnetic field is no longer dipolar, and a transition into the active high-multiplicity magnetosphere occurs. To narrow down the scope for a look into finer details, we concentrate on the single photon conversion process, since the thermal $X$-ray photons required for the two-photon interactions (serving as the abundant targets for the less numerous curvature $\gamma$-ray photons to hit) would unlikely be sufficiently populous, because the stellar temperature of around $10^6$K \cite{2016JPlPh..82e6302P} would be too low \cite{2002A&A...384..414P} (the two-photon process can however be important for more powerful pulsars when current sheets are present, see next section for more discussions). 

For aligned and orthogonal rotators, the site for pair cascading by the single photon absorption process would likely be substantially divergent. With orthogonal rotators, if $\Delta\Phi^{\rm vac}$ is large enough, pair cascade could occur far away beyond the trap surfaces, since the rotation induced electric field as given by \eqref{eq:EthField} and \eqref{eq:EphField} contains slowly declining $1/r^2$ pieces when $\alpha = \pi/2$. In contrast, the leading contribution to the electric field is $1/r^4$ (quadrupolar) for aligned rotators. Indeed, computation by \cite{2002A&A...384..414P} shows that this process is likely operative only very close to the star, within around tens of stellar radii. One possible site for aligned rotator pair cascade is then in the crevice between the dome and the disk (close to the orange thread between the red and blue clouds of Fig.~\ref{fig:Dead}(a)). Such a case is indeed seen in the axisymmetric simulations by \cite{2019arXiv191100059C}, where an intermediate $\Delta\Phi^{\rm vac}$ corresponds to a cyclic behavior of the magnetosphere, switching between activity and comatose, bearing obvious resemblance to the nulling behavior. The electronics analogy of such a magnetosphere is perhaps an LC circuit, with a capacitor in the form of a charge cloud residing at the \verb!Y! point, as opposed to an open circuit as electrospheres or properly closed circuit as the active case. 

Although not yet numerically investigated, we can expect that if we further relax axisymmetry, another cascade site would become available. The reason being that the diocotron/slipping-stream instability (see Appendix \ref{sec:QuantitativeChargeSeparate}) would become activated in this case, causing the equatorial disk to develop into high charge density vortices in the azimuthal direction, which subsequently merge and either collapse to hug the stellar surface closely, turning the entire large radii regions vacuum, or, when particle injection is introduced, form complex radially-extended patterns containing prominent vacuum gaps (see Figs.~9 and 11 of \cite{2009A&A...503....1P}). These gaps cannot be closed up by the pairs produced within (they are not closed even by artificially injected single-signed charges), since the walls bounding a gap are all made up of the same charges. Specifically if, say, electrons produced by pair production, re-enforce and thus draw one wall in, the positrons of the pairs would inevitably annihilate and push back the border on the other side. These long-lasting gaps, residing within the disk where the magnetic field is strong and single photon conversion is efficient, would then keep injecting charges into the magnetosphere, turning the pulsar on (see e.g., \cite{2014ApJ...785L..33P,2018ApJ...858...81B} for active magnetospheres sustained by particle injected from close to the star). Furthermore, the diocotron instability grows on timescales comparable to pulsar rotation periods \cite{2002A&A...387..520P,2007A&A...464..135P,2009A&A...503....1P}, implying that the transition from ``off'' to ``on'' states can be accomplished within timescales as short as a few pulse intervals, which would indeed be necessary if it is to explain the nulling behavior. Furthermore, the cyclic behavior resulting from pair cascading only in the ``crevice'' is also of periods comparable to rotation period \cite{2019arXiv191100059C}, thus either site (``crevice'' or ``disk opening''), or indeed both, can be relevant for nulling.  

For a more quantitative assessment, we note that the approximate criteria for single photon cascading is given by \cite{1966RvMP...38..626E} (see their Eq.~1.4 and discussion below Eq.~1.5, also their Fig.~12) as 
\bea
\chi\equiv \frac{\epsilon_{\gamma}}{2m_e c^2}\frac{B}{B_{\rm cr}} \gtrsim 10^{-1}\,,
\eea
where $\epsilon_{\gamma}$ is the characteristic energy associated with the physical process under consideration (photon energy in our context), $B$ is the magnetic field strength and 
\bea
B_{\rm cr} \equiv \frac{m^2_e c^3}{e\hbar}= 4.414\times 10^{13}\text{G}\,.
\eea
The electron rest energy on the other hand is 
\bea
m_e c^2 = 0.51\text{MeV} = 8.2\times 10^{-7} \text{erg}\,.
\eea
When applied to magnetospheres, we would want nearly full conversion of photons as they traverse the gap. Let $\ell_{\rm s}$ be the full conversion shielding distance, then Eq.~3.6 of \cite{1966RvMP...38..626E} yield the requirements [these expressions are derived under the assumption $B\ll B_{\rm cr}$, which is reasonably valid for typical pulsar magnetospheres of $\mathcal{O}(10^{12})$G]
\bea
B \ell_{\rm s} \sim 4.6\,, \quad \frac{\epsilon_{\gamma}}{m_e c^2} \gtrsim 6\times 10^7 \ell_{\rm s}\,,
\eea
where $B$ is in units of MG, and $\ell_{\rm s}$ is in cm. For an active gap with characteristic dimension $\ell_{\rm g}$, we would need  $\ell_{\rm s} \lesssim \ell_{\rm g}$, we have thus the conditions 
\bea \label{eq:SinglePhotonConversion}
\mathfrak{C}^e_1:\ell_{\rm g} \gtrsim \frac{4.6}{B}\,, \quad 
\mathfrak{C}^e_2:\frac{E}{m_e c^2} \gtrsim 6\times 10^7\,,
\eea
where we have used $\epsilon_{\gamma} \sim E \ell_{\rm g}$ (we are considering the minimal requirement that there exists sufficiently energetic photons, these photons may not be the most populous among those produced by synchrotron radiation), with $E$ being the electric field strength in the gap with a unit of statV/cm. The condition $\mathfrak{C}^e_1$ is quite simple to satisfy since neutron star vicinity has typical length scales of kilometers, so $\ell_{\rm g} \sim 10^5$cm, thus $B \sim 100$G would already be sufficient, meaning that $\mathfrak{C}^e_1$ is satisfied for regular pulsars all the way out to the light cylinder. The second condition $\mathfrak{C}^e_2$ is $E\sim 50$ statV/cm, which is a little more stringent. From vacuum approximations for which we have analytical estimates \eqref{eq:ErField}-\eqref{eq:EphField}, we get that 
\bea
E_{\rm align} \sim \frac{B_0 \Omega }{2 c}\frac{R^5}{r^4}\,, \quad 
E_{\rm ortho} \sim \frac{B_0 \Omega }{2 c}\frac{R^3}{r^2}\,. 
\eea   
Substituting in the typical parameters $B_0 \sim 10^{12}$G, $\Omega \sim 1$, and $R\sim 6$km, we have that the maximum distances at which we could still have sufficient $E$ strength are at around $r=127$km and $r=2700$km for the aligned and orthogonal rotators respectively. 

These numbers are in agreement with our naive expectation that the electrosphere-destroying pair cascades can occur far outside of the trap surfaces for the orthogonal rotators, but can only occur quite close to the star for the aligned rotator. Although we won't be able to predict the exact location of the transition lines due to the limitations imposed by the various approximations we have to adopt (chiefly that the field strength expressions are not those of actual electrospheres), the fact that these numbers, computed with realistic pulsar parameters, are greater than neutron star radius and smaller than the light cylinder radius, suggests that the electrospheres should not be uniformly stable nor unstable for the entire regular pulsar population (if both numbers are smaller than stellar radius, we will not have any place with a sufficiently strong $E$ to break down the electrospheres; and if both numbers are larger than typical light cylinder radius, then there will always be gaps powerful enough to destroy electrospheres, so this quiescent configuration can never exist in reality) -- it is indeed plausible for there to be two sub-populations possessing very different magnetospheric structures.  

\subsection{Replenishing an active magnetosphere} \label{sec:Active}
The quiescent charge-separated electrospheres 
has a muted discharge with all the plasma supplied by the star, which is sufficient since there is no pulsar wind and thus no voracious demand for replenishment. 
The high multiplicity active magnetosphere on the other hand, must be underpinned by an active pair cascade inside gaps where the electric field parallel to the magnetic field is not shorted out. These gaps are formed afresh when the magnetosphere turns active and are not necessarily associated with the gaps in the electrospheres. 
Once such gaps open up, the same single-photon conversion consideration of the previous section applies, but there is now the prerequisite that such gaps would not be quickly shorted out by the surrounding plasma (as opposed to the electrosphere case whereby stable gaps are known to always be present). We turn now to this additional condition, on the stable existence of the necessary gaps that continuously supply plasma to maintain a charge neutral force-free magnetosphere -- if the gaps cannot persist, the unbalanced loss of charged particles through pulsar winds will cause the magnetosphere to become lower multiplicity, evolving towards the charge separated electrospheres. 

Previous studies (see e.g., \cite{2013MNRAS.429...20T,2014ApJ...795L..22C,2015ApJ...801L..19P,2015ApJ...815L..19P}) have found that the main criteria for efficient pair production in the polar cap is that $\mathfrak{C}^a$: the local current density $j_{\parallel}$ along the magnetic field lines should exceed the Goldreich-Julian or GJ value 
\bea \label{eq:GJCurrent}
j_{\rm GJ}=c\rho_{\rm GJ} = -2\epsilon_0 c \frac{{ \Omega} \cdot { B}}{1-\left|\frac{{ \Omega} \times { r}}{c}\right|^2}\,, 
\eea
so that the particle number density as inferred from $j_{\parallel}$ still exceeds the GJ { density} value\footnote{A charge deficiency on the other hand would not lead to a lapse in shielding, as it only causes a charge-density wave to set up but the mean flow is still the GJ flow \cite{1997MNRAS.287..262S}.}, even if the charges are moving with a maxed-out speed $v\sim c$ (thus the charge density requirement is at a minimum). Such over-charging prevents the proper shielding of the $E_{\parallel}$ field, instead resulting in an(other) $E_{\parallel}\neq 0$ gap region that keeps accelerating charged particles while they move away from the star, and the photons emitted via curvature radiation or energized via inverse-Compton scattering can trigger pair cascade via single photon conversion \cite{2014MNRAS.441.1879P}. 
The potential difference that develops in such a gap is approximated by \cite{2008ApJ...683L..41B} 
\bea \label{eq:GapPotential}
\Delta\Phi \sim 4\pi\mathcal{R}\left(\zeta-1 \right)\frac{c \rho_{\rm GJ}}{c}h^2  \,,
\eea
topping out at $\Delta \Phi^{\rm vac}$, where $h < r_{\rm pc}$ is the height of the gap ($r_{\rm pc}$ being the radius of the polar cap), and $\mathcal{R}$ is the Ramp function.

It is worth noting that many studies { (e.g., \cite{1985A&A...149...57L})} would assume force-freeness \emph{a priori} for mathematical tractability (i.e., within the formalism adopted, already assuming that the charge density is at the GJ value and $E_{\parallel}$ is shielded), then condition $\mathfrak{C}^a$ is equivalently stated for them as there being regions where the 4-D current $J^a$ becomes spacelike $J^aJ_a>0$ (because the temporal component $J^0=c\rho$, equalling $c\rho_{\rm GJ}$ by the force-free assumption, is too small as compared to the spatial component $j_{\parallel}$, exactly the same situation that $\mathfrak{C}^a$ describes), flagging a consistency problem (charges need to move superluminally to produce such a current, which is impossible) indicating that the force-free assumption breaks down, and we would instead witness a gap there. 
Also worth noting is the fact that we have thus far described only a single-charge-species simplification serving as a time-averaged toy model. The full picture is more complex and dynamic, with the charged particle number density possibly many times that of the GJ value (but since both charges are present, average charge density can be much lower), and with pair production likely occurring in bursts (counter-streaming opposite charges can momentarily keep $j_{\parallel}$ high while retaining charge density at $\rho_{\rm GJ}$ in localized regions, but cannot maintain such a balance everywhere in a stationary manner, see \cite{2013MNRAS.429...20T,2014MNRAS.441.1879P,2015ApJ...815L..19P}; so spacelike currents still flag the need for pair production, but the temporal and spatial features could be much richer). 

There is furthermore an alternative site for pair production relevant for young and fast spinning pulsars \cite{1986ApJ...300..500C,2014ApJ...795L..22C}, in the current sheets where $\zeta\equiv j_{\parallel}/j_{\rm GJ}<0$ \cite{2014MNRAS.441.1879P}, so that the charges have the wrong sign for shielding the electric field. The accelerated charges can emit synchrotron photons [e.g., $\gamma$-ray signals from pulsars can be produced this way \cite{2015MNRAS.448..606C}], and some of which can then produce pair cascade via $\gamma-\gamma$ interactions. The demands on the magnetic field strength is relaxed as compared to the single photon conversion, since the magnetic field does not participate directly in this process, which can then occur in the current sheets further from the star. However, hints are that the polar caps are likely the more important pair production sites when the pair-injection rate is low \cite{2018ApJ...858...81B} (meaning lower overall pair flux is required if injected at stellar surface only), which is the relevant regime for our present considerations. Indeed, computations based on single photon magnetic conversion predicts a pair creation death valley at $\Delta \Phi^{\rm vac} \approx 10^{12}$V \cite{Ruderman:1975ju}, which matches the observed pulsar radio emission turn-off region \cite{2009ASSL..357..373A,2014MNRAS.441.1879P}. 

The determination of $j_{\parallel}$ is nontrivial however, since if we see the magnetosphere as a circuit, then some resistivity comes from the crossing of the twisted magnetic field lines by the particles on the outside \cite{1997MNRAS.287..262S}, where the interaction with interstellar medium is complicated. 
Nevertheless, it appears reasonable to expect that: 
\begin{enumerate}
\item 
The aligned (and anti-aligned) rotators are not effective accelerators, because $j_{\rm GJ}$ is quite large when ${ \Omega}$ and ${ B}$ are aligned, making it more difficult for $j_{\parallel}$ to exceed it. This leads to the likelihood that the previously derived $\mathscr{P}_{\mathcal{A}}\geq \mathscr{P}_{\mathcal{A}}^e$ reduces to a necessary but not sufficient condition, namely that while it being satisfied is sufficient for pair cascade if there is a gap, the gap won't in fact be available at this $\mathscr{P}_{\mathcal{A}}$ yet. A more restrictive $\mathscr{P}\geq \mathscr{P}_{\mathcal{A}}^a$ fulfilling Eq.~\eqref{eq:Stripe} is needed instead. 

Evidences for this behavior exist in literature, e.g., analytical models predict $\zeta < 1$ \cite{2006MNRAS.368.1055T}, and numerical simulations such as those done by \cite{2014ApJ...795L..22C} and \cite{2015ApJ...801L..19P} confirmed that for nearly aligned and anti-aligned rotators, charge-separated electrosphere solutions are obtained (for the non-juvenile pulsars). Such a complete shut-down for all low inclination pulsars can however be remedied by General Relativistic frame-dragging effects \cite{2015ApJ...815L..19P}, which reduce the effective angular velocity of the star and subsequently $j_{\rm GJ}$. This fact would be consistent with the $\mathcal{A}$ stripe being not far from the overall pulsar death line. 

\item
With orthogonal rotators, the story is different. Because ${ \Omega}$ and ${ B}$ are nearly orthogonal, $j_{\rm GJ}$ nearly vanishes and condition $\mathfrak{C}^a$ is easily satisfied. However, this also translates into the fact that we are working with much smaller numbers and the resulting accelerating electric field may not be sufficiently large to drive pair cascades. Indeed, the indicative vacuum electric field as given by Eq.~\eqref{eq:ErField} with $\alpha=\pi/2$ and $\theta=0$ gives a vanishing time-averaged $\langle E_r \rangle$ in the polar regions. Alternatively, substituting Eq.~\eqref{eq:GJCurrent} into Eq.~\eqref{eq:GapPotential}, we see that since ${ \Omega} \cdot { B} \approx 0$ for an orthogonal rotator, the electric field inside the gap would be $E\approx \Delta \Phi/\ell_{\rm g} \approx 0$. In other words, $\zeta >1$ only ensures that the electric field won't be shielded, it does not guarantee that the resulting unshielded field is strong enough to accelerate charges to energies sufficient for pair production, as demanded by Eq.~\eqref{eq:SinglePhotonConversion}\footnote{This subtler extremal case has not been scrutinized by numerical studies of generic oblique rotators, which are more interested in identifying all possible potential sites of active pair formation of a generic oblique rotator. In particular, numerical simulations tend to pick an unrealistically (at least for non-millisecond pulsars) large ratio $\xi=R/R_{\rm LC}$ between stellar and light cylinder ($R_{\rm LC}$) radii, in order to avoid having to simultaneously handle two drastically different scales, meaning small time steps are needed to resolve the short scale but the simulation has to run for a long time (thus an enormous number of evolution steps) to see any changes in the slow scale ($1/\xi$ is roughly the number of time steps needed). This in effect massively boost the rotation angular speed of the star, which is $\propto 1/R_{\rm LC}$, thus boosts $\Delta \Phi^{\rm vac}$ and masks the potential problem with orthogonal rotators. For example, \cite{2015ApJ...801L..19P} adopted a value of $\xi= 0.3$ (in contrast, for a realistic regular pulsar with a period on the order of a second, $\xi \sim 10^{-5}$), and as such did not see any issue with pair cascade with orthogonal rotators. }.

As a result, the existence of the polar gaps is not restrictive and the opening up of the band (where high fractional nullers are found) as required by Eq.~\eqref{eq:Stripe} is not due to the extra requirement $\mathfrak{C}^a$. Instead, the fact that $\mathscr{P}_{\mathcal{O}}^a$ exceeds $\mathscr{P}_{\mathcal{O}}^e$ is simply due to the polar gaps of active magnetospheres residing close to the magnetic axis, where the rotation-induced electric field is suppressed, while the gaps of the electrosphere occupy other latitudes where the electric field is much larger.

\end{enumerate}

\subsection{Intermittency} \label{sec:Intermittent}
Beyond the magnetospheric changes that we have considered so far, there is a further complication relating to evolutions within the neutron star. Namely, that the magnetic field configurations
within the star can contain higher order multipoles, which will affect the field structure immediately abutting the star (but not further away, since the higher order multipoles drop off much more quickly with radial distance), and significantly boost pair cascade. It does so by two means: ($\mathfrak{e}_1$) the polar cap shifts location, which turns the effective inclination less extreme; ($\mathfrak{e}_2$) the curvature of the magnetic field lines are enhanced so curvature radiation becomes more efficient. 

If this happens, an otherwise borderline ``off'' extreme-inclination pulsar may be turned ``on''. For moderate-inclination pulsars, similar internal magnetic field dynamics would also be present, but with less dramatic impacts -- the effective inclination is already non-extremal, so altering it a little won't impart significant qualitative changes to pair cascading. The two extreme-inclination cases also differ significantly, as the following simple argument already shows: referring to Fig.~\ref{fig:Vacuum} for a crude dipolar illustration, we see that with an orthogonal rotator (second row in the figure), the Lorentz force $-(e/c){ v}_{e}\times { B}$ due to the star's rotation, as experienced by those highly mobile electrons within the neutron star near the dipolar cap region, is pointed along the $z$ axis (with the direction being consistent for all electrons near a polar cap), meaning that the electrons and thus currents sustaining the magnetic field will be pushed around as compared to the lowest order dipole approximation (thus offsetting from the positively charged immobile ions, producing an electric field, to achieve force balance as per usual Hall effect), so the centers of the polar caps will likely shift from being in the equatorial plane to somewhere with a nonvanishing $z$ value; in contrast, this force is axisymmetric against the $z$ axis for an aligned rotator, so the polar caps' centers will remain on that axis, and the polar caps will thus only change in size, but will not shift in location. In other words, effect ($\mathfrak{e}_1$) is suffered more acutely by orthogonal rotators.

Furthermore, if there are multiple equilibrium states that are similar in energy, but some with greater higher order multipoles giving rise to abundant pair production and thus active magnetospheres, and some less, leading to subdued pair production and thus quiescent magnetospheres, then we would observe the internal field and current configuration switching between these states on Hall evolution timescales\footnote{The Hall drifting is the advection of magnetic field by free electrons, which is the most important dynamics \cite{2014PhRvL.112q1101G} in the crust consisting of a neutral fluid with ions pinned down. Ohmic dissipation can also occur but conductivity is quite high even in the crust so it only operates on much longer timescales for pulsars of regular $\mathcal{O}(10^{12})$G magnetic field \cite{2013arXiv1305.0149G} -- essentially all previous studies on the crustal magnetic field concentrate on pulsar magnetic field decay over thousands to millions of years, but even for them, Ohmic decay is subdominant (but can be non-negligible). It is nevertheless interesting to note that hypothetically, should Ohmic decay becomes important on our timescale, it would tend to dissipate higher order multipoles \cite{2014PhRvL.112q1101G}, and the ``on'' state would become less viable. Finally, ambipolar diffusion is expected to occur deeper down where more particle species are mobile, but, as per common industry practice, we adopt the simplest Meissner condition that the superconducting core expels the magnetic field so only the crust is important for us. This is justified because the strength of ambipolar diffusion declines as magnetic field strength cubed, so is expected to be important only for young magnetars \cite{2019arXiv191103095P}.}, driving the magnetosphere to hop between active and comatose states. 
In other words, we have intermittency for orthogonal rotators (more pedestrian radio emission mode-changing may be due to the same physics, but for moderately oblique and aligned rotators), which is a different thing from nulling, since the pulsar jumps between truly (quasi-stationary) ``on'' and ``off'' states, rather than being stuck in the middle band between $\mathscr{P}_{\mathcal{O}}^a$ and $\mathscr{P}_{\mathcal{O}}^e$ (although the ``on'' state for some cases could obviously also land in the middle zone). 

Turning to more details, we have the Hall evolution equation  
\bea \label{eq:EOM}
\frac{\partial  {B}}{\partial t} = -\frac{c}{4\pi e n_e} {\nabla} \times \bigg[\left({\nabla} \times { B}\right)\times { B}\bigg]\,,
\eea
which is simply the Faraday induction law supplemented by the Hall-drifting condition on the particles (i.e., ignoring the non-Hall terms in the generalized Ohms law for plasmas, such as pressure gradient and the small resistivity \cite{Spitzer1962}; see \cite{1997ITPS...25.1229B} for the physical conditions that the negligence of those terms corresponds to)
\bea \label{eq:Drift}
{ E} = -\frac{{ v}_e\times { B}}{c}\,,
\eea
as well as the neglecting of the displacement current (thanks to the high conductivity in the star) in the Amp\`ere's law so ${ v}_e \propto { j} \propto \nabla\times { B}$, where ${ v}_e$ is the electron velocity and ${ j}$ is the current density. 
Eq.~\eqref{eq:EOM} is nonlinear so we won't be able to solve it to provide quantitative descriptions. Nevertheless, it has the following qualitative/semi-quantitative properties that argue for its relevance to the intermittency phenomenon:
\begin{enumerate}
\item Because the Hall drift conserves energy, there is no reason (on the short timescales we are interested in, when Ohmic dissipation is negligible) for the internal magnetic structure to be stuck in some equilibrium state. It would instead be able to wonder around along equi-energy surfaces in the space of such structures (i.e., the shifted polar caps will likely move about). 

Specifically, Eq.~\eqref{eq:EOM} is hyperbolic, with component form resembling the well-known Burgers' equation (after introducing simplifying symmetries) \cite{2019arXiv191103095P}. As such, it generically exhibit wave-like periodic behavior -- manifesting as the low frequency helicons/whistlers \cite{Aigrain1961,1969AmJPh..37..241M} in the linearized limit. Such oscillatory behavior is indeed seen in the numerical simulations of e.g., \cite{1997A&A...321..685S} (although the star is simplified into a uniform one rather than having a crust-core division for this study, and the setup is axisymmetric). 

Note that while the underlying internal magnetic field structure evolves continuously, the external pair cascade should turn on and off in a more clear-cut binary manner, because the relevant microphysics is characterized by sharper threshold conditions, such as $\mathfrak{C}^e_2$. 

\item Note that when deriving Eq.~\eqref{eq:EOM}, we have neglected inertial forces such as the Coriolis force that would be seen in the co-rotating frame, since they are negligible in strength (contributing to the force balance at equivalent to around only $200$G of magnetic field strength). This amounts to the zero electron mass limit (the strong magnetic field means that the cyclotron radius is small and so the microscopic cyclotron motion of the electrons won't complicate the dynamics), which removes the Trivelpiece-Gould mode and disables the associated potential enhancement on plasma production at the conductor surface. As a consequence, the motion of the electrons is determined completely by the electromagnetic field, and so just as with the force-free case, we have that the particle motions contributing to a nonlinear modification to an otherwise vacuum electromagnetism. 

Unlike most discussions in helicon literature (see introductory texts such as \cite{1969AmJPh..37..241M}), our physical setup is not one where we can approximate the magnetic field as a small perturbation on top of a constant background. In other words, we cannot linearize the problem, and the periodicity in our nonlinear setting cannot be expected to be very regular (indeed, the rigorous mathematical definition of ``waves'' in the nonlinear context is subtle). This nonlinearity would generically cause the evolution of the internal magnetic field to appear rather stochastic. 

Note for theoretical studies such as \cite{1997A&A...321..685S} though, one has to impose artificial symmetries for tractability, thus will see more organized periodic behavior. In other words, one should expect observed intermittency to exhibit less clean periodicity than seen naively in theoretical predictions. Observationally, the intermittent pulsars are indeed seen to be quasi-periodic, see e.g., \cite{2013ApJ...775...47C}. 

\item 
The Hall drifting timescale is given by \cite{1992ApJ...395..250G} as (robust and not sensitive to the state of matter in the neutron star; also dimensionally consistent with Eq.~\ref{eq:Drift})
\bea \label{eq:HallTime}
t_{\rm Hall} 
\approx 5\times 10^8 \, \frac{L^2_{\rm km}}{B_{\rm 12}}\frac{\rho}{\rho_{\rm nucl}}\, \text{yr}\,,	
\eea
where the characteristic magnetic field strength $B_{\rm 12} = B/10^{12}\text{G} \sim \mathcal{O}(1)$ for regular pulsars, while $L_{\rm km}=L/10^5\text{cm}$ is the length scale at which the magnetic field density changes, which is again $\sim \mathcal{O}(10^{-1})$ to $\mathcal{O}(1)$, being limited from above by the thickness of the neutron star crust, and below by the size of the polar cap (we are mostly interested in the lowest few higher order multipoles that would change the overall structure of the polar caps if ($\mathfrak{e}_1$) is the most important effect; if ($\mathfrak{e}_2$) is more important, higher order multipoles may also become relevant). The constant $\rho_{\rm nucl}=2.8\times 10^{14}$ $\text{g}/\text{cm}^3$ is the nuclear density, and $\rho$ is the density of the neutron star in the upper crustal region of interest, which is around $8.0 \times 10^{6}$ $\text{g}/\text{cm}^3$ [specifically, the density of ${}^{56}\text{Fe}$ in the ground state of cold dense matter\footnote{Note simulation works studying magnetic field decays don't normally include this solid upper crust (see e.g., \cite{2014MNRAS.438.1618G}; simulations are also restricted to the axisymmetric case), because it introduces much smaller timescales than the one they are interested in, and thus requires too many time steps, but it is precisely such short timescales that are relevant for our intermittency modelling effort.} \cite{2006PhRvC..73c5804R}]. Substituting all these numbers into Eq.~\eqref{eq:HallTime}, we get $t_{\rm Hall}$ on the order of a few months or years, in good agreement with observed intermittency timescales. 

This $t_{\rm Hall}$ being quite long is a result of the helicon waves being much more slowly moving than vacuum electromagnetic waves. We essentially have a helicon standing wave in the crust, the phase of which the higher order multipoles of the magnetic field depends on. Our range of $t_{\rm Hall}$ obtained above is also consistent with the oscillation period seen in the numerical simulations of \cite{1997A&A...321..685S}. 
\end{enumerate}

\section{Discussion and conclusion} \label{sec:Discussion}
We highlighted an apparent feature seen in the distribution of the high-fraction nulling and intermittent pulsars on the $\log_{10}P-\log_{10}\dot{P}$ diagram, namely their congregating near two stripes of constant polar electric potential. We analysed the statistical significance of this feature and proposed an explanation. 

Our discussions are exploratory. In particular, we caution that due to limited amounts of availability data (in e.g., the number of intermittent pulsars, the polarization measurements etc), which are further contaminated by the complication of the inclination angle not being perfectly $0$ or $\pi/2$, we do not claim that the existence of the claimed stripes is firmly established. We aim instead to motivate further observational efforts towards their confirmation or rejection, ideally involving observational campaigns on the relevant pulsars, using highly sensitive radio telescopes that would be able to see weak interpulses. Such observations have the potential to directly verify the extreme inclination hypothesis.  

The theory part of the note are also only semi-quantitative, due to the fact that we are discussing finer details in the pulsar behavior, when consensus on even the more rudimentary pulsar emission mechanism is yet elusive. We have thus to evoke various subtitles that must combine to provide us with the observations, while each still remaining an active debate within the pulsar theory community. Under such circumstances, firm conclusions are not prudent. Nevertheless, simple order of magnitude estimates appear to offer up reasonable matches with observations, so we have adventured to discuss the explanation, with the aim to offer an example of how the aforementioned observational campaigns, if carried out, could benefit the theoretical modeling efforts.  

Lastly, we wish to clarify that we have concentrated only on the high-fraction nullers because the low-fraction ones may just be due to our sight line missing the carousel of sub-beams, or the extinguishing of some sub-beams \cite{1970Natur.228...42B,1976MNRAS.176..249R,1983AIPC..101..163A,2001MNRAS.322..438D,2008MNRAS.385.1923R,2009MNRAS.395.1529R} (the formation of such beams may be quite dynamic and thus capricious, see e.g., \cite{2018A&A...612A..24Z} for an example), and are thus less mysterious. In particular, observational hints exist that point to differences in the underlying mechanisms driving low and high-fraction nulling. For example, the debate as to whether nulling occurs simultaneously across frequencies could possibly be resolved if one notes that studies yielding an affirmative result appear to have examined high fraction nullers [e.g., $75\%$ for B0826-34 \cite{2008MNRAS.383.1538B}], while those answering in the negative looked at lower fraction ones instead [e.g., $1.4\%$ for B0809+74 \cite{1978A&A....70..307B}, $15\%$ for B1133+16 \cite{2007A&A...462..257B}]. 
Indeed, our proposed existence of two separate bands, for only the high-fraction nullers\footnote{The Rotating Radio Transients could also be very high fraction nullers, but they could just as well be due to the giant pulses of young pulsars that rise above detection threshold \cite{2003ApJ...596..982M}. Since we cannot really distinguish between the various possibilities [unfortunately, they cannot all be nullers, or else the population synthesis doesn't work out \cite{2006Natur.439..817M,2013IAUS..291...95B}], we exclude this class of objects from our considerations in this note and include only confirmed nulling pulsars.}, if true, represents a somewhat complicated structure and could possibly help explain the apparently conflicting conclusions reached by different studies (see e.g., \cite{1976MNRAS.176..249R,1986ApJ...301..901R,1992ApJ...394..574B,2007MNRAS.377.1383W}) regarding correlations among parameters of nulling pulsars. That is, there is indeed an underlying structure, so correlations do exist, yet different relationships are obeyed by separate sub-populations, thus varying conclusions can be drawn when studying different samples.  

\section*{Acknowledgements}
We would like to thank Duncan Lorimer for bringing to our attention the existence of the $\mathcal{O}$ stripe that intermittent pulsars congregate around. This work is supported by the National Natural Science Foundation of China grants 12073005, 12021003, 11503003 and 11633001, the Interdiscipline Research Funds of Beijing Normal University, and the Strategic Priority Research Program of the Chinese Academy of Sciences Grant No. XDB23000000.

 \bibliographystyle{ieeetr}
\bibliography{../../../../../References}


\appendix
\onecolumn

\section{Models of pulsar magnetospheres} \label{sec:MagSpheres}
\subsection{Vacuum magnetosphere} \label{sec:Lines}
In a pulsar magnetosphere, the vacuum dipole induced potential drop $\Delta\Phi^{\rm vac}$ across the polar caps (the base region of the field lines that reach beyond the light cylinder) is always a useful baseline parameter, providing an estimate on the typical energy of charged particles lifted out of the star that might be available to seed cascades, regardless of whether the polar cap is the actual site of cascading (such as proposed by \cite{1970Natur.227..465S,1971ApJ...164..529S,Ruderman:1975ju}). For example, even with young pulsars where cascading happens predominantly in the current sheets, a transition of the magnetosphere into a charged-starved electrosphere state is observed by \cite{2014ApJ...795L..22C} to occur when the threshold photon energy for pair production becomes comparable to $e\Delta\Phi^{\rm vac}$.

Following the treatment of \cite{Goldreich:1969sb,2012hpa..book.....L} for an aligned rotator, we have that inside of the neutron star seen as a perfect conductor, the charged particles must not experience any net force, so 
\bea \label{eq:ForceFree}
{ E}_{\rm ind} + \frac{1}{c}\left( { \Omega} \times { r}\right)\times { B} = 0,
\eea 
where ${ E}_{\rm ind}$ is an electric field produced by charge redistribution in the conductor (it is not an induction field generated by the variation of the magnetic field; e.g., for an aligned dipole, we have $\partial { B}/\partial t=0$ but this ${ E}_{\rm ind}$ does not vanish). Substituting in the expression for a dipolar magnetic field, one can solve for the vacuum electric scalar potential outside of the star due to the induced charge redistribution, yielding the quadrupolar 
\bea \label{eq:Pot}
\Phi^{\rm vac} = -\frac{B_0 \Omega R^5}{6cr^3}\left(3\cos^2\theta -1\right)\,,  
\eea 
where $R$ is the stellar radius, $B_0$ is the magnetic field strength at the equator, and $\theta$ is measured against the rotation axis. One can then further evaluate the potential drop across the polar cap, which would be responsible for lifting charges off of the star to serve as seeds for pair production cascades ($e^{\pm}$ discharge). To this end, one begins by demarcating the boundary of the polar cap, by noting that the boundary field lines will touch the light cylinder tangentially at the equatorial plane (given the reflection symmetry for an aligned rotator), or $\theta=\pi/2$. Since $\sin^2\theta/r$ is a constant along a dipolar magnetic field line, we have that the angle $\theta_P$ at which the boundary line strikes the stellar surface is given (for aligned rotators) by
\bea
\sin^2\theta_P = \frac{R\Omega}{c} \sin^2\frac{\pi}{2}\,. 
\eea 
Substituting into Eq.~\eqref{eq:Pot}, we then have that 
\begin{align}  \label{eq:PotDiffAlign}
\Delta \Phi^{\rm vac} \equiv& \Phi^{\rm vac}_{\theta=\theta_P} - \Phi^{\rm vac}_{\theta=0} 
=\frac{B_0 \Omega R^5}{6cr^3}\left(3\sin^2\theta_p\right)
= \frac{B_0 \Omega^2 R^3}{2c^2}\,. 
\end{align} 
One then further replace $B_0$ by its estimate from dipole radiation (only roughly valid if dipole radiation contributes a roughly constant proportion of energy loss for any given fixed inclination angle $\alpha$)
\bea \label{eq:DipoleB}
B_0 = \sqrt{\frac{3c^3}{8\pi^2}\frac{I}{R^6\sin^2\alpha}P\dot{P}}\,,  
\eea 
where $I\approx (2/5)MR^2$ is the neutron star moment of inertia ($M$ is stellar mass), resulting in lines of 
\begin{align} \label{eq:Line}
-\frac{3}{2}\log_{10}P+\frac{1}{2}\log_{10}\dot{P} \equiv \mathscr{P}(\alpha)
\end{align}
marking out constant $\Delta \Phi^{\rm vac}$ values in the $\log_{10}P-\log_{10}\dot{P}$ diagram when $\alpha$ is held constant at a small value (because the treatment leading up to Eq.~\eqref{eq:PotDiffAlign} is valid only for aligned rotators; we have also assumed that the relative stellar mass and radius variation between pulsars is less than order unity, as variations in $\sin\alpha$ are). For this case, we further have explicitly that  
\bea \label{eq:LineSmallAngle}
\mathscr{P} = \log_{10}\left(\frac{\sin\alpha \sqrt{5}\sqrt{c}\Delta \Phi^{\rm vac}}{\sqrt{3}\pi\sqrt{M} R}\right)\,.
\eea

\begin{figure*}[ptb]
  \centering
\begin{overpic}[width=0.32\textwidth]{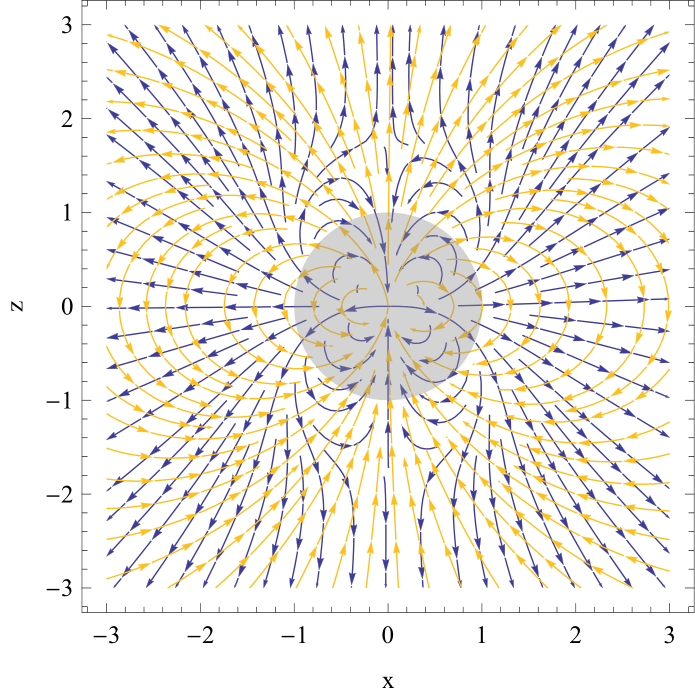}
\put(5,2){(a)}
\end{overpic}
\begin{overpic}[width=0.32\textwidth]{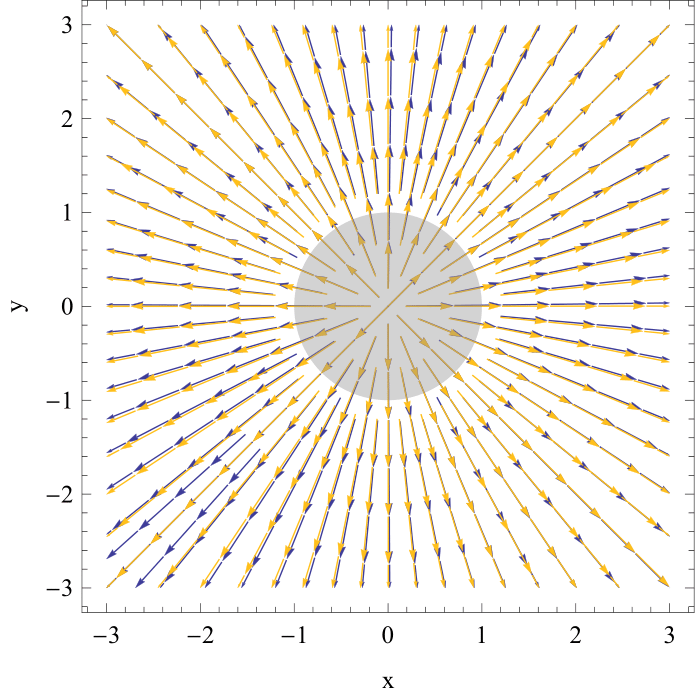}
\put(5,2){(b)}
\end{overpic}
\begin{overpic}[width=0.32\textwidth]{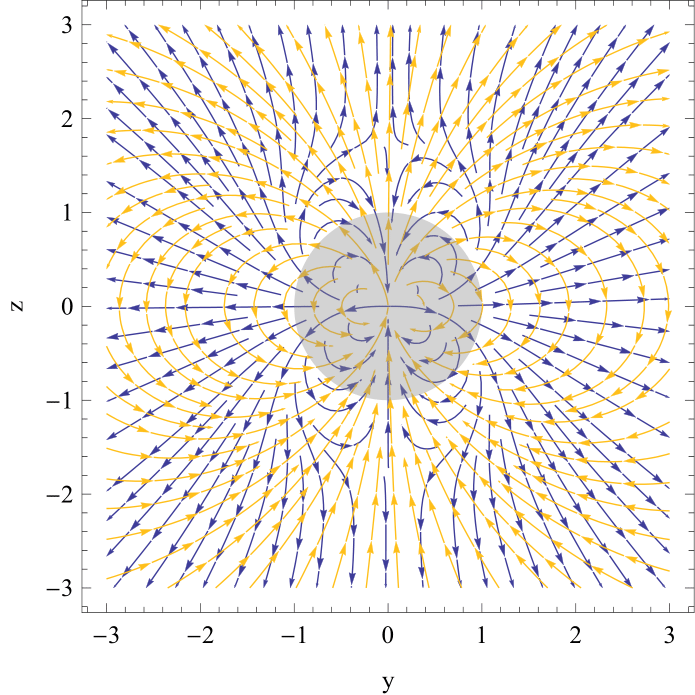}
\put(5,2){(c)}
\end{overpic}
\begin{overpic}[width=0.32\textwidth]{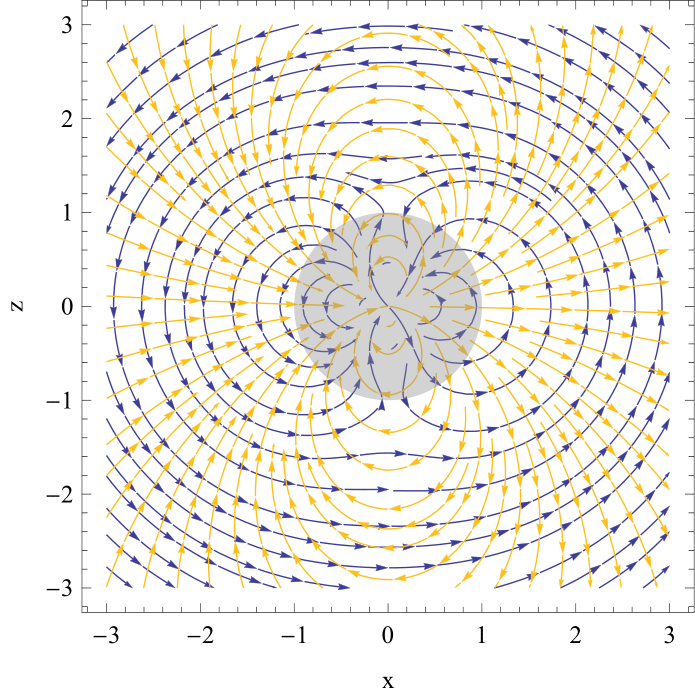}
\put(5,2){(d)}
\end{overpic}
\begin{overpic}[width=0.32\textwidth]{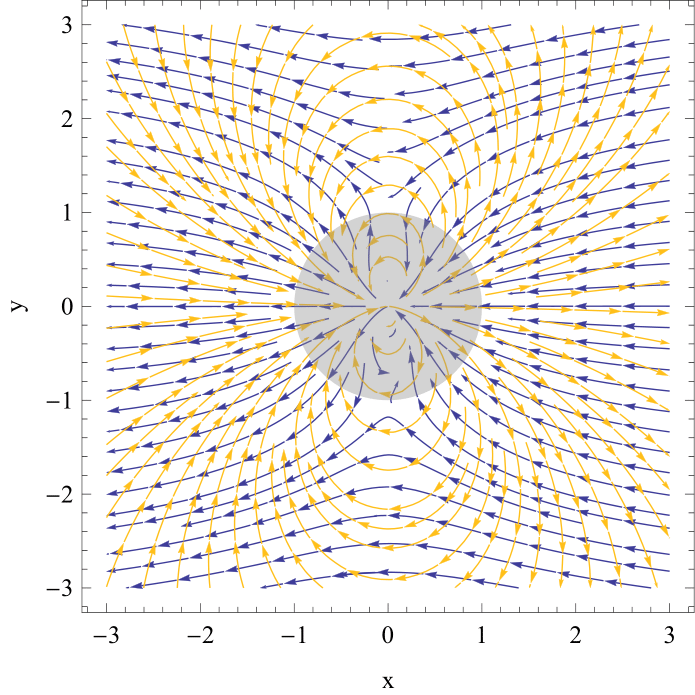}
\put(5,2){(e)}
\end{overpic}
\begin{overpic}[width=0.32\textwidth]{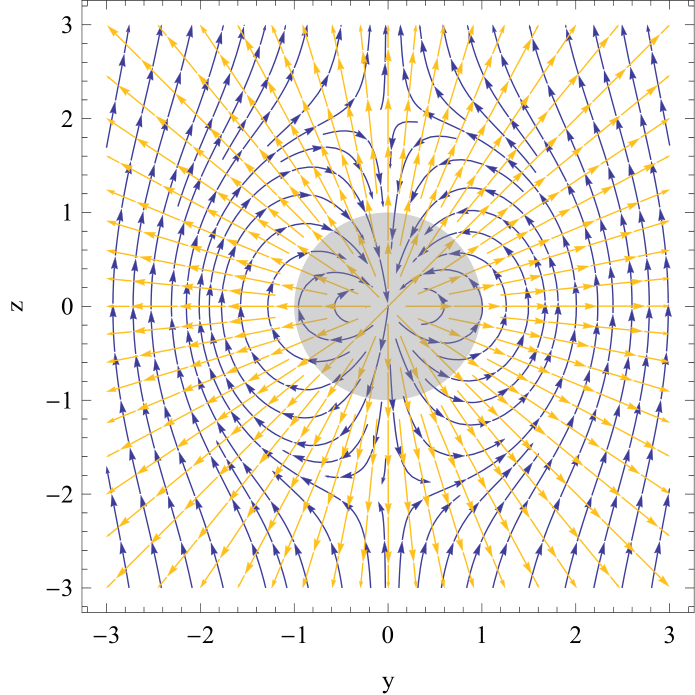}
\put(5,2){(f)}
\end{overpic}
\begin{overpic}[width=0.32\textwidth]{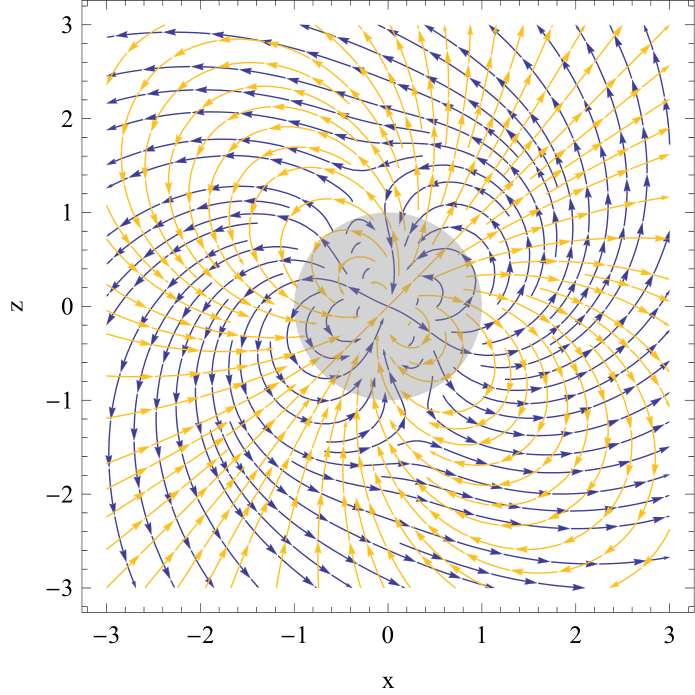}
\put(5,2){(g)}
\end{overpic}
\begin{overpic}[width=0.32\textwidth]{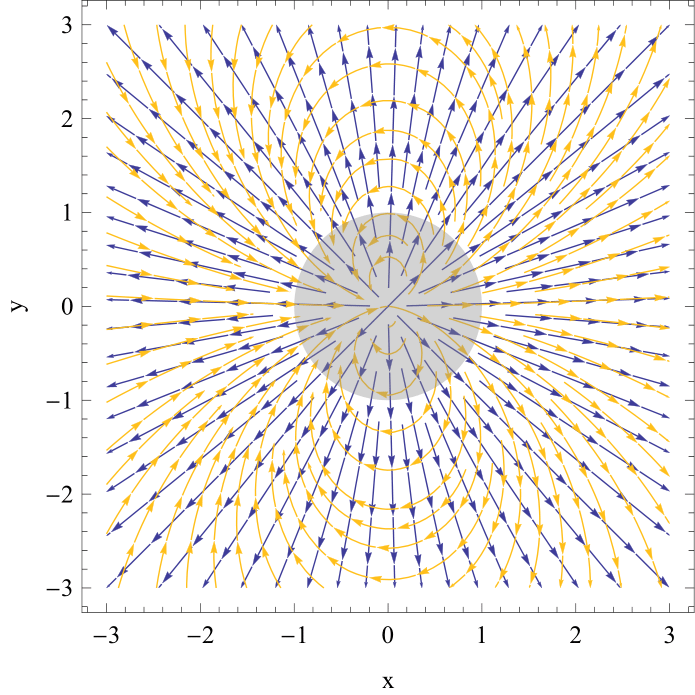}
\put(5,2){(h)}
\end{overpic}
\begin{overpic}[width=0.32\textwidth]{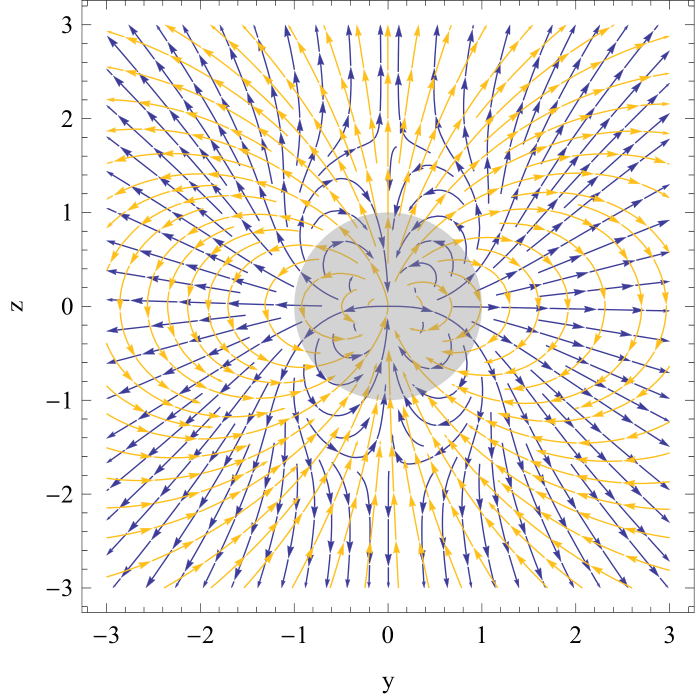}
\put(5,2){(i)}
\end{overpic}
  \caption{The magnetic (yellow, light in black and white) and electric (blue, dark) field lines projected onto the $x-z$ (first column), $x-y$ (second column), and $y-z$ (third column) planes, for the aligned (first row, the rotation axis is in the $z$ direction, and the magnetic axis is in the $x-z$ plane for the moment plotted), orthogonal (second row) and moderately oblique ($\alpha=\pi/4$; third row) rotators. The semi-transparent disk represents the stellar surface, and note that the plotting software produces spoke patterns when the field on the plane is vanishing. 
  }
	\label{fig:Vacuum}
\end{figure*}

This treatment can also be generalized to oblique rotators, 
e.g., \cite{1955AnAp...18....1D,1999PhR...318..227M,2016JPlPh..82e6302P} give (the vector basis $\{{ e}_{r},{ e}_{\theta},{ e}_{\phi}\}$ is orthonormal)
\begin{align} 
B_r =& \frac{B_0 R^3}{r^3} \bigg( \cos\alpha\cos\theta+\sin\alpha\sin\theta\cos(\varphi-\Omega t)\bigg)\,,\label{eq:BrField}  \\
B_{\theta} =& \frac{B_0 R^3}{2r^3} \bigg( \cos\alpha\sin\theta-\sin\alpha\cos\theta\cos(\varphi-\Omega t)\bigg)\,,\label{eq:BthField}  \\
B_{\varphi} =& \frac{B_0 R^3}{2r^3} \sin\alpha\sin(\varphi-\Omega t)\,,\label{eq:BphField} \\
E_r =& -\frac{B_0 \Omega R^5}{2 c r^4} \Bigg[ \cos\alpha \left(3\cos^2\theta-1\right)
+\frac{3}{2}\sin\alpha \sin2\theta \cos(\varphi-\Omega t)\Bigg]+\frac{Q}{4\pi\epsilon_0 r^2}\,, \label{eq:ErField} \\ 
E_{\theta} =& \frac{B_0 \Omega R^3}{2 c r^2} \Bigg[ \sin\alpha \left(\frac{R^2}{r^2}\cos2\theta-1\right)\cos(\varphi-\Omega t)
-\frac{R^2}{r^2}\cos\alpha \sin2\theta \Bigg]\,, \label{eq:EthField} \\
E_{\varphi} =& \frac{B_0 \Omega R^3}{2 c r^2}\left(1-\frac{R^2}{r^2}\right) \sin\alpha \cos \theta \sin(\varphi-\Omega t)  \label{eq:EphField} \,, 
\end{align}
where $Q$ is the overall charge on the star, which is zero for either dipolar or uniform magnetizations inside of the star. The expressions above are visualized in Fig.~\ref{fig:Vacuum}, and they ignore the displacement current $\partial_t { E}$, so back-action on the magnetic field is neglected. This will be a common theme encountered throughout neutron star literature and this paper. It is justified for the near-star region of regular pulsars for which the rotation period is on the order of seconds, so we have $R\Omega/c\sim 10^{-5} \ll 1$ (more intuitively manifesting as the light cylinder being very far away from the star), thus $|{ E}|\ll |{ B}|$. 
The displacement current expression in the full Amp\`ere-Maxwell equation 
\bea \label{eq:MaxwellAmpere}
\nabla \times { B} = \frac{1}{c} \frac{\partial { E}}{\partial t} + \frac{4\pi}{c} { j}
\eea 
further brings in another such factor ($\partial_t$ introduces the $|{ v}|$ dependence), so the back-action is only very mildly perturbative ($\nabla \times$ gives a factor of $\mathcal{O}(10^{-6})\text{cm}^{-1}$ suppression to the left hand side for even the global neutron star scale features, thus the ratio of the perturbations to the original magnetic field is around $\mathcal{O}(10^{-4})$). One should note though that the situation is different with millisecond pulsars. 

Integrating $E_r$ along $r$ direction to infinity then gives $\Phi^{\rm vac}$ equaling $(r/3)E_r$. The second line in Eq.~\eqref{eq:ErField} averages to zero over a rotation period thus does not contribute to a time-average $\langle \Delta \Phi^{\rm vac} \rangle $. For an oblique rotator, the potential difference
\begin{align} \label{eq:PotDiff}
\langle \Delta \Phi^{\rm vac} \rangle =& \langle \Phi^{\rm vac} \rangle_{\theta=\theta_{P1}}-\langle \Phi^{\rm vac} \rangle_{\theta=\theta_{P2}} 
= \frac{B_0 \Omega R^5}{6cr^3}\left(\sin^2\theta_{P1}-\sin^2\theta_{P2}\right)
\end{align}
between the two ends of the polar cap that differ the greatest in $\theta$ (since $\langle\Phi^{\rm vac}\rangle$ is $\theta$ dependent) is what's interesting to us (in contrast, for an aligned rotator, the rim of the polar cap share the same $\theta_P$, which was why we used the $\theta =0$ reference point to evaluate the maximum potential drop). To evaluate $\theta_{P1/2}$, we note that the magnetic field outside of the star is essentially a superposition of aligned and orthogonal rotators, including and in particular $B_r$ and $B_{\theta}$ as given by Eqs.~\eqref{eq:BrField} and \eqref{eq:BthField}. 
Assume that a boundary field line strikes the light cylinder tangentially at a location with angle $\theta_{\rm LC}$, and thus
\bea
\tan\theta_{\rm LC} = \frac{B_{\theta}}{B_r}\,, 
\eea 
where the right hand side is independent of $r$, we have then that $\theta_{\rm LC}$ must be independent of how far the light cylinder is to the star, or in other words independent of $\Omega$. We thus have that 
\bea \label{eq:Ang}
\sin^2\left(\theta_{P1/2}-\alpha\right) = \frac{R\Omega}{c} F_{1/2}\,. 
\eea 
where $F_{1/2}$ depends on $\alpha$ but not $\Omega$. For $\alpha=\pi/2$, the left hand side of Eq.~\eqref{eq:Ang} becomes $\cos^2\theta_{P1/2}$ which can be substituted directly into Eq.~\eqref{eq:PotDiff} resulting in once again $\langle \Delta \Phi^{\rm vac} \rangle \propto \Omega^2 B_0$, and subsequently Eq.~\eqref{eq:Line} remains valid. 

It is important to remember that the right hand side of Eq.~\eqref{eq:Line} is $\alpha$ dependent, so the lines are only meaningful when comparing pulsars with similar inclinations. Therefore, the observed huddling of intermittent and high-fraction nulling pulsars near the $\mathcal{O}$ and $\mathcal{A}$ stripes (each has a slope consistent with Eq.~\ref{eq:Line}) is indicative that they should be grouped into two particular inclinations. Furthermore, at a given $P$ and $\dot{P}$, $B_0$ for larger inclination angles will be smaller, so $\Delta \Phi^{\rm vac}$ at that point in the $\log_{10}P-\log_{10}\dot{P}$ diagram will be smaller for larger inclinations. Alternatively and more intuitively, \cite{Goldreich:1969sb} computed the induced charge density inside of the star to be $-2\epsilon_{0}{ \Omega} \cdot { B}_0 = -2\epsilon_{0}\Omega B_0 \cos\alpha $, so the associated electric field is smaller for larger inclinations. In turn, if there is a roughly common shut-down threshold $\Delta \Phi^{\rm vac}$ value for both nearly-aligned and orthogonal pulsars, one would expect the orthogonal death line to occur further to the top-left of the diagram than the aligned death line, which is indeed consistent with the placement of the $\mathcal{O}$ and $\mathcal{A}$ stripes. 

Finally, we note that the vacuum treatment of this section provides us with only the maximum possible $\mathscr{P}$ that would be available. In reality, since we are interested in charge-filled magnetospheres, it is not the entire $\Delta\Phi^{\rm vac}$, but its charge-modified version that manages or fails to drive the cascades. We thus turn to plasma filled electrospheres.

\subsection{With plasma: the active magnetospheres} \label{sec:QuantitativeActive}
\begin{figure}[tb]
  \centering
\begin{overpic}[width=0.375\columnwidth]{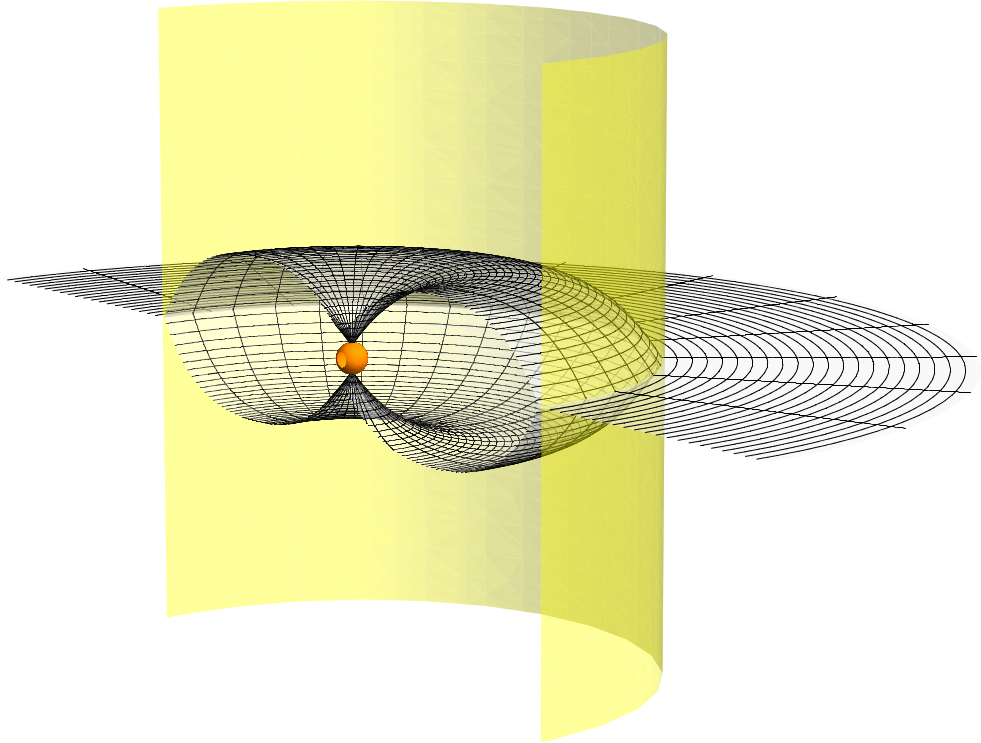}
\end{overpic}
  \caption{The structural backbone of an active magnetosphere for an aligned rotator (sliced in half). The semi-transparent yellow surface is the light cylinder, beyond which plasma cannot co-rotate with the star without going superluminal. The white meshed surfaces are the current sheets, with the two pieces within the light cylinder enclose the closed field lines region in which the plasma particles co-rotate. The field lines residing outside will penetrate the light cylinder and sweep back in the azimuthal direction, along which the particles stream (thus not co-rotating with the star). These out-flowing pulsar wind particles could cross the magnetic field lines in the outer regions where the magnetic field strength is weak, and return through the current sheets, completing a full circuit that maintains charge neutrality. The neutron star is depicted as an orange sphere at the center, whose size is, for illustrative purposes, massively enlarged from what should be for common pulsars. 
  \vspace{5mm}
  }
	\label{fig:Alive}
\end{figure}

Although \cite{Goldreich:1969sb} (GJ) already painted a semi-quantitative picture of a plasma-filled magnetosphere, the precise details of the charge distribution and generation mechanism has been the subject of extensive debate. Earlier analytical and numerical work { \cite{1976ApJ...206..831J,1985A&A...149...57L,1985MNRAS.213P..43K,1985A&A...144...72K,2001MNRAS.322..209S} }showed that charges lifted from the star itself will form force-free charge-separated ``electrosphere'' structures (crudely stylized in Fig.~\ref{fig:Dead}(a)(b) below) that are rather different from the GJ prescription (see Sec.~\ref{sec:QuantitativeChargeSeparate} for more details). 
These solutions are well constructed but predicts inactive magnetospheres, as they are characterized by charge trapping [thus no current/pulsar wind flowing out to the nebula \cite{1976ApJ...206..831J}]. Interests in them thus waned and the search went on to look for active magnetospheres.

Later on, numerous studies (e.g., \cite{1999ApJ...511..351C,2014ApJ...785L..33P,2015ApJ...801L..19P,1974ApJ...193..225P,2009ApJ...690...13M}) successfully showed that the GJ-like configuration is instead achieved when, in addition to charges from the star, abundant $e^{\pm}$ pair production \cite{1971ApJ...164..529S,1976ApJ...203..209C} by cascading 
is \emph{assumed}, turning the magnetosphere into a high multiplicity (neutral plasma) one. In these solutions, currents flow out along the open field lines, and return through the current sheets (styled in Fig.~\ref{fig:Alive}, see also e.g., \cite{2016ApJ...833..258G} and references therein), forming an active circuit that maintains charge neutrality, thereby answering a fundamental critique leveled against the original GJ model. This is the canonical pulsar magnetosphere that has nowadays become widely accepted. We do not summarize much further details about its properties, since the review literature on it is extensive and easy to find. We do emphasize however, that the magnetic field structure for the active magnetosphere is dipolar only in the near zone (potentially containing higher multipoles immediately abutting the star), and becomes split monopolar at far away due to the strong impact of the high current density within the current sheets. On the other hand, the magnetic field for the electrospheres that we describe next do not differ much from the vacuum case of Eqs.~\eqref{eq:BrField}-\eqref{eq:BphField} (the electrical field become shielded within charge clouds, and are no longer described by Eqs.~\ref{eq:ErField}-\ref{eq:EphField}, but since the back-reaction on the magnetic field is small, this change in the electric field does not alter the magnetic field). 

\subsection{With plasma: the charge-separated electrospheres} \label{sec:QuantitativeChargeSeparate}

\begin{figure}[tb]
  \centering
\begin{overpic}[width=0.22\columnwidth]{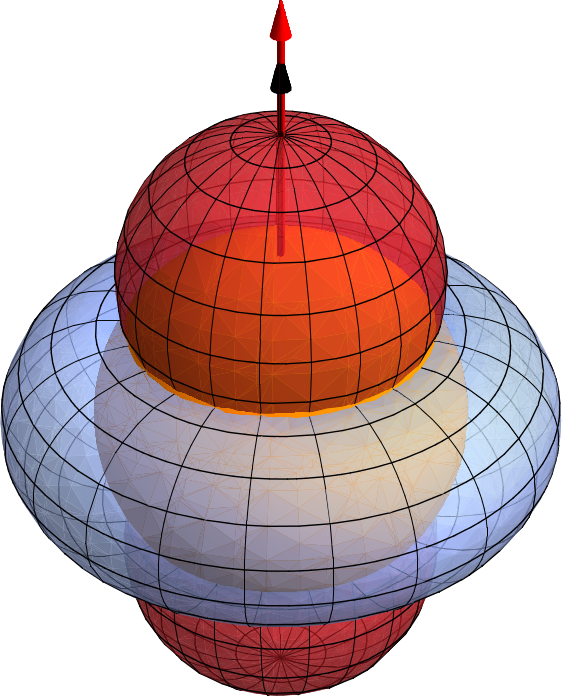}
\put(5,2){(a)}
\end{overpic}
\begin{overpic}[width=0.27\columnwidth]{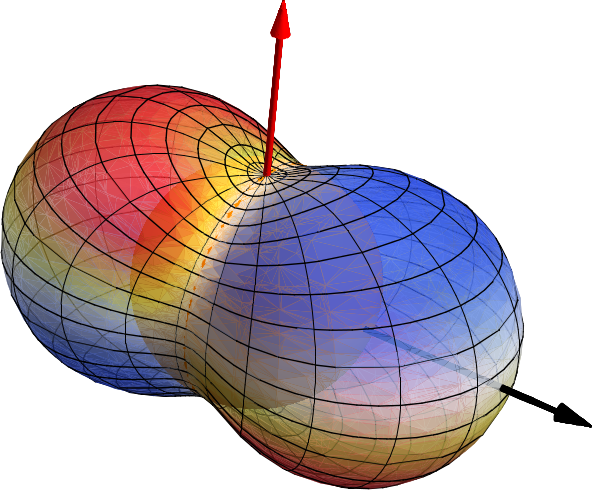}
\put(5,2){(b)}
\vspace{3mm}
\end{overpic}
\begin{overpic}[width=0.20\columnwidth]{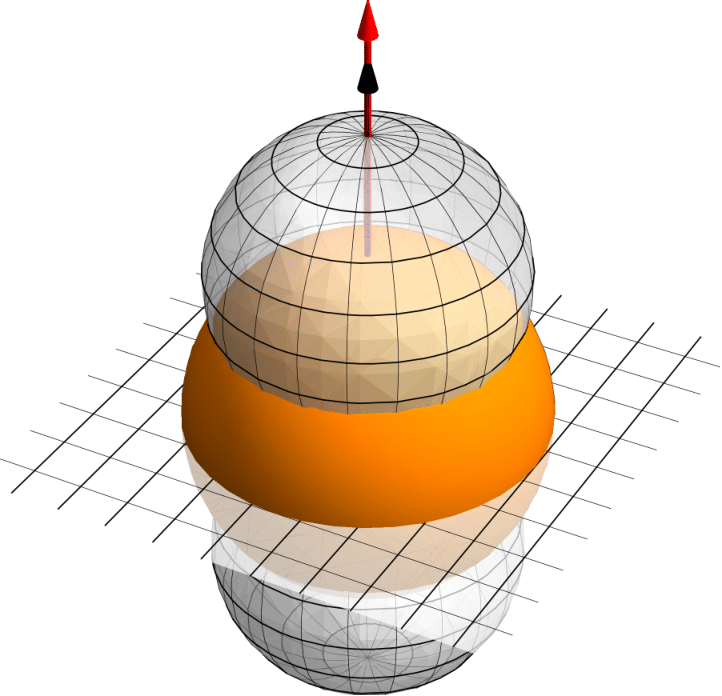}
\put(5,2){(c)}
\end{overpic}
\begin{overpic}[width=0.27\columnwidth]{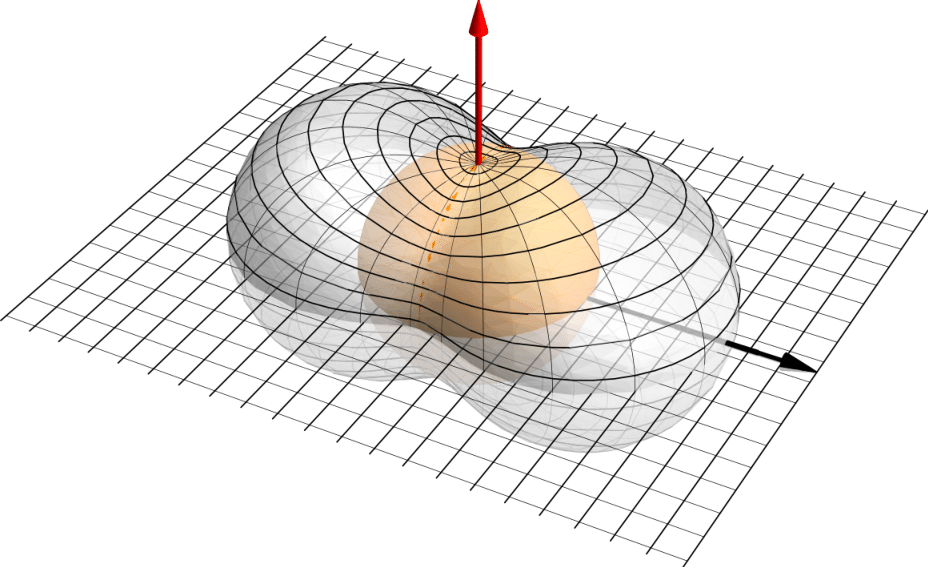}
\put(5,2){(d)}
\end{overpic}
  \caption{Schematic drawings of the electrospheres. (a) Electrosphere for aligned rotators, resembling the rotating Terrela configuration familiar to plasma physicists. (b) Electrosphere for orthogonal rotators. The meshed semi-transparent surfaces are the sharp boundaries of the charged regions. The plasmas with charges indicated by the color (red for negative, blue for positive) fill the space between these surfaces and the stellar surface (the orange surface faintly visible behind the meshed surface). There are no currents flowing in any region of the electrospheres.  
 (c) (d) The trap surfaces in vacuum magnetospheres as given by Eq.~\eqref{eq:Trap}. The orange sphere signifies the neutron star, the red arrow is the rotation axis and the black arrow the magnetic axis. 
  \vspace{5mm}
  }
	\label{fig:Dead}
\end{figure}

We summarize now the aforementioned charge-separated electrospheres. These are of a dome plus torus structure for aligned rotators \cite{1976ApJ...206..831J,1985MNRAS.213P..43K,1985A&A...144...72K,2001MNRAS.322..209S}, and quadrants of alternating charges for orthogonal rotators \cite{1974ApJ...193..225P,2009ApJ...690...13M} (see Fig.~\ref{fig:Dead}). More quantitatively, it was noted by \cite{2009ApJ...690...13M} that the sharp boundaries of the charge distribution tracks closely the vacuum solution's force-free trapping surfaces, i.e., places where $E_{\parallel} =0$ is already satisfied in the vacuum solution \eqref{eq:BrField}-\eqref{eq:EphField} (and continuity dictates that $E_{\parallel}$ is of opposite signs on either side of it). Such surfaces are natural discontinuous boundaries between charge clouds and vacuum gaps. On the force-free side, the charges ensure that $E_{\parallel}=0$ (and if any lapse in shielding occurs, the electric field direction is such that it drives charges towards the trap surface, which is why the cloud fills at least up to this surface), but on the vacuum side, $E_{\parallel}$ is non-vanishing and points in the expedient direction to collect and transport stray particles of the same (sign of) charge into the force-free side, while repelling particles of the undesirable opposite charge. This ensures that the charge clouds, and in turn the electrosphere solutions, are stable \cite{1985A&A...144...72K,2005A&A...429..779A}\footnote{It is to be noted that some literature claims that the GJ magnetosphere is unstable and would collapse into an electrosphere. The GJ magnetosphere studied there is not the modern version with current sheets though. The plasma examined was also not of high multiplicity/nearly-neutral. Nevertheless, they demonstrate the stability of the electrospheres under non-drastic (no turning-on of new sources of particles via pair cascade) perturbations. In any case, this mechanism only works for charge-separated, not neutral plasma \cite{2001MNRAS.322..209S}, and is therefore irrelevant for the active magnetospheres}. Such a self-maintenance feature also prevents clouds from dissipating if they spill over beyond the trap surfaces, as would be the case when there is too much charge and the electric repulsion is severe. In other words, the trap surfaces mark only the minimal sizes of the clouds, but do not set the upper bound. 

Explicitly, these trapping surfaces are (derived from vacuum -- the actual boundaries of the charge clouds are deformed from these due to the presence of charges, but this higher order correction does not admit a simple analytical description)
\begin{eqnarray}\label{eq:Trap}
  \left\{
  \begin{aligned}
&\theta=\frac{\pi}{2} \,\, \text{or} \,\, r=\sqrt{3} R |\cos \theta|\,,  \,\, \text{for} \,\, \alpha = 0\,,\\
&\theta=\frac{\pi}{2} \,\, \text{or} \,\, r=R \sqrt{2-\cos 2\theta+2 \cos2\phi \sin^2\theta}\,, \,\, \text{for} \,\, \alpha = \frac{\pi}{2}\,,
  \end{aligned}
  \right.
\end{eqnarray}
and plotted in Fig.~\ref{fig:Dead}(c)(d). 

For aligned rotators, we can further write out the expression 
\bea
{ E}\cdot { B} = -4 R^2 \cos \theta  (3 \cos 2 \theta +1)\,,
\eea
which changes sign at 
\bea
\theta^{\rm c} =(1/2)\cos ^{-1}\left(-1/3\right)\approx 55^\circ\,, 
\eea
meaning that charges of opposite signs are lifted out of the star in regions $\theta < \theta^{\rm c}$ and $\theta > \theta^{\rm c}$, so that $\theta^c$ is the angle at which the trap surface intersects the star. When combined with 
the trapping surfaces of Eq.~\eqref{eq:Trap}, and the fact that charges can only move along magnetic field lines, one then arrives at the following (leading order) qualitative expectation: charges of one sign (negative if the rotation and magnetic axes are aligned, positive if anti-aligned) are lifted from the polar ($\theta < \theta^{\rm c}$) regions of the star and moved to the dome-like structures in Fig.~\ref{fig:Dead}(c) along the polar magnetic field lines; charges of the opposite sign are lifted from the equatorial ($\theta > \theta^{\rm c}$) regions of the star and channelled to the equatorial plane along those dipolar magnetic field lines striking the star at $\theta > \theta^{\rm c}$. Because $\sin^2\theta/r$ is conserved along the dipolar field lines, there is a maximum radial extend (where the magnetic field line hitting the star at $\theta^{\rm c}$ crosses the equatorial plane)
\bea
r^{\rm max} = R \frac{\sin^2\frac{\pi}{2}}{\sin^2\theta^{\rm c}}= \frac{3}{2}R\,,
\eea
that the latter charges can reach along the equatorial plane. In other words, these charges do not permeate the entire plane in Fig.~\ref{fig:Dead}(c), forming instead only a torus quite close to the star as shown in Fig.~\ref{fig:Dead}(a). More specifically, one can envision that the torus is bounded by the aforementioned field line with shape $r=(3R/2)\sin^2\theta$. It is observed that the charge density in the torus and dome are similar to the GJ values \cite{2001MNRAS.322..209S}. Such a configuration is depicted in Fig.~\ref{fig:Dead} (a), but is only a leading order approximation. In particular, while the domes co-rotate with the star, and the part of the equatorial thick disk/torus abutting the star is in co-rotation, the further away regions are in differential (even faster) rotation \cite{2002A&A...384..414P} (the differential rotation is needed to evade the restrictions laid down by \cite{1974ApJ...190..391P}, which only applies to axisymmetry, so no such complication needs to arise for the orthogonal rotator). The differentially rotating disk is shown to suffer from diocotron instability \cite{2002A&A...387..520P,2007A&A...464..135P,2014ApJ...785L..33P}, thus exhibits non-axisymmetric charge modulations and radial expansion \cite{2014ApJ...785L..33P}, and the actual equatorial charge distribution likely is more complicated than just laminar flow around a simple torus, with macroparticles/vortices likely showing up \cite{2009A&A...503....1P} [also note that the disk is not bound by a trap surface, so there is no aforementioned self-maintenance mechanism at work to regulate the shape of the disk; nevertheless, the disk remains confined, at least in the absence of pair injection, and cannot escape to infinity \cite{2005A&A...434..405A}].  

For the orthogonal rotators on the other hand, \cite{2009ApJ...690...13M} noted that the region within the peanut-shaped force-free trapping surfaces agree roughly with the analytical prescription given by \cite{1974ApJ...193..225P} (derived assuming a particular form for the current, among other simplifications), with the charge density being (note unlike below, the original Eq.~31 in the reference did not explicitly write out the $\phi$ dependent factors, but its Eq.~25 correspond to $l=1$ and solves Eq.~23 only when $m=\pm 1$; also it was in SI units, which we translate into cgs)
\begin{align} \label{eq:OrthoPeanut}
\rho_{\rm peanut}=&-\frac{3\Omega B_0}{4\pi c}\frac{R^3}{r^3}\sin 2\theta
\left[1-\frac{\Omega^2 r^2}{6c^2}\left(1-9\sin^2\theta\right)\right]\cos\phi\,,
\end{align} 
but confined to regions much closer to the stellar surface, rather than reaching outwards to regions close to the light cylinder as this equation would predict (see Fig.~\ref{fig:Dead}(b)).  
The corresponding nonvanishing current density component is 
\begin{align}
j_{\rm peanut \, \phi}=&-\frac{3\Omega^2 B_0}{4\pi c}\frac{R^3}{r^2}\sin\theta\sin 2\theta
\left[1-\frac{\Omega^2 r^2}{6c^2}\left(1-9\sin^2\theta\right)\right]\cos\phi\,. 
\end{align}
Similar to the electric field in Eqs.~\eqref{eq:ErField}-\eqref{eq:EphField}, this current is also of the order $R\Omega/c \sim \mathcal{O}(10^{-5})$ times $B_0$ near the stellar surface, but without the benefit of another $R\Omega$ supplementing it in Eq.~\eqref{eq:MaxwellAmpere}, so its back-action on the magnetic field is even smaller than the displacement current, thus also negligible. 

\end{document}